\documentclass[twocolumn, tighten, times]{aastex61}
\usepackage{graphicx}
\usepackage{epstopdf}
\usepackage{lineno}

\usepackage{threeparttable}
\usepackage{sidecap}
\sidecaptionvpos{figure}{c}
\usepackage{floatrow}
\shorttitle{PGC 44685: A Dwarf Star-forming Lenticular Galaxy with Wolf-Rayet Population}
\shortauthors{Lu et al. }
\begin{document}

\title{PGC 44685: A Dwarf Star-forming Lenticular Galaxy with Wolf-Rayet Population}

\correspondingauthor{Qiusheng Gu}
\email{qsgu@nju.edu.cn}

\author{Shiying Lu}
\affil{School of Astronomy and Space Science, Nanjing University, Nanjing, Jiangsu 210093, China }
\affil{Key Laboratory of Modern Astronomy and Astrophysics (Nanjing University), Ministry of Education, Nanjing 210093, China}

\author{Qiusheng Gu}
\affil{School of Astronomy and Space Science, Nanjing University, Nanjing, Jiangsu 210093, China }
\affil{Key Laboratory of Modern Astronomy and Astrophysics (Nanjing University), Ministry of Education, Nanjing 210093, China}

\author{Yulong Gao}
\affil{School of Astronomy and Space Science, Nanjing University, Nanjing, Jiangsu 210093, China }
\affil{Key Laboratory of Modern Astronomy and Astrophysics (Nanjing University), Ministry of Education, Nanjing 210093, China}

\author{Yong Shi}
\affil{School of Astronomy and Space Science, Nanjing University, Nanjing, Jiangsu 210093, China }
\affil{Key Laboratory of Modern Astronomy and Astrophysics (Nanjing University), Ministry of Education, Nanjing 210093, China}

\author{Luwenjia Zhou}
\affil{School of Astronomy and Space Science, Nanjing University, Nanjing, Jiangsu 210093, China }
\affil{Key Laboratory of Modern Astronomy and Astrophysics (Nanjing University), Ministry of Education, Nanjing 210093, China}

\author{Rub{\'e}n Garc{\'i}a-Benito}
\affil{Instituto de Astrof{\'i}sica de Andaluc{\'i}a-CSIC, Glorieta de la Astronom{\'i}a s.n., Granada, 18008, Spain}

\author{Xiangdong Li}
\affil{School of Astronomy and Space Science, Nanjing University, Nanjing, Jiangsu 210093, China }
\affil{Key Laboratory of Modern Astronomy and Astrophysics (Nanjing University), Ministry of Education, Nanjing 210093, China}

\author{Jiantong Cui}
\affil{School of Astronomy and Space Science, Nanjing University, Nanjing, Jiangsu 210093, China }
\affil{Key Laboratory of Modern Astronomy and Astrophysics (Nanjing University), Ministry of Education, Nanjing 210093, China}

\author{Xin Li}
\affil{School of Astronomy and Space Science, Nanjing University, Nanjing, Jiangsu 210093, China }
\affil{Key Laboratory of Modern Astronomy and Astrophysics (Nanjing University), Ministry of Education, Nanjing 210093, China}

\author{Liuze Long}
\affil{School of Astronomy and Space Science, Nanjing University, Nanjing, Jiangsu 210093, China }
\affil{Key Laboratory of Modern Astronomy and Astrophysics (Nanjing University), Ministry of Education, Nanjing 210093, China}

\author{Zhengyi Chen}
\affil{School of Astronomy and Space Science, Nanjing University, Nanjing, Jiangsu 210093, China }
\affil{Key Laboratory of Modern Astronomy and Astrophysics (Nanjing University), Ministry of Education, Nanjing 210093, China}

\begin{abstract}
Lenticular galaxies (S0s) are formed mainly from the gas stripping of spirals in the cluster. But how S0s form and evolve in the field is still untangled. Based on spatially resolved observations from the optical Hispanic Astronomical Center in Andalusia 3.5-m telescope with the PPAK Integral Field Spectroscopy instrument and NOrthern Extended Millimeter Array, we study a dwarf ($\rm M_*<10^9 M_\odot$) S0, PGC 44685, with triple star-forming regions in the central region, namely A, B, and C, respectively. In northwest region C, we clearly detect the spectral features of Wolf-Rayet (WR) stars and quantify the WR population by stacking spectra with high WR significance. Most of the molecular gas is concentrated in the region C(WR), and there is diffuse gas around regions A and B. The WR region possesses the strongest intensities of H$\alpha$, CO(1-0), and 3mm continuum, indicating its ongoing violent star formation (gas depletion timescale $\lesssim$25 Myr) with tentative hundreds ($<$500) km/s stellar winds accompanied by the WR phase. Most ($\sim 96\%$) of three star-forming regions show relatively low metallicity distributions, suggesting possible (minor) accretions of metal-poor gas that trigger the subsequent complex star formation in a field S0 galaxy. We speculate that PGC 44685 will become quiescent in less than 30 Myr if there is no new molecular gas to provide raw materials for star formation. The existence of this dwarf star-forming S0 presents an example of star formation in the low-mass/metallicity S0 galaxy.

\end{abstract}


\section{Introduction}
There is a notable dichotomy in several aspects between spiral and elliptical galaxies, which is evident in the bimodal distributions of characteristics such as the presence or absence of spiral arms, blue or red color, and rotational or pressure-dominated kinematics, etc (\citealt{Mishra+17}). A ``more or less hypothetical'' transitional stage between spirals and ellipticals was then introduced as lenticular galaxies (i.e., S0s) by \cite{Hubble+1926}, which features a prominent spheroidal component surrounded by a stellar disk. The most important channel for forming S0s is a ``faded spiral'' quenching scenario by the consumption or removal of gas from blue star-forming spirals (\citealt{Barr+07, Bekki+11}), compared to the ``merger'' scenario (\citealt{Mendez-Abreu+18, Coccato+22}) or the internal secular evolution by some morphological features such as bars (\citealt{Laurikainen+06, Fraser-McKelvie+18a}). Thanks to two-dimensional (2D) integral field unit (IFU) survey studies, such as Mapping Nearby Galaxies at APO (MaNGA), the radial distribution of stellar population for individual galaxy could be spatially resolved in the literature, such as \cite{Fraser-McKelvie+18b} and \cite{Dominguez-Sanchez+20}.  They found a mass-based bimodality of S0s that massive ($\rm > 10^{10} M_\odot$) S0s have older bulges than disks, resulting from inside-out quenching by the ``merger'' scenario. In comparison, less massive ($\rm < 10^{10} M_\odot$) S0s tend to go through the ``faded spiral'' scenario to produce younger bulges.

Like the majority of early-type galaxies (ETGs),  most normal S0s contain little gas and show a low level of star formation (e.g., \citealt{Welch+10, Xiao+16, Xu+22}). But nuclear star-forming activities are found in some S0s, called star-forming S0s (SFS0s). \cite{Xiao+16} found 8\% (45) SFS0s in the total (583) S0s sample, mainly based on the Sloan Digital Sky Survey (SDSS) Data Release 7 (DR7). \cite{Xu+22} found 52 SFS0s compared to a normal (216) S0s sample, based on the SDSS-IV MaNGA survey. Both studies found that SFS0s have smaller stellar mass ($\rm M_*$) than normal S0s, and the measurement of the S{\'e}rsic index ($n$) hints at the existence of pseudo-bulges (average $n\sim$1.6). As more and more observations detected atomic or/and molecular gases in S0s (\citealt{Welch+10}) and SFS0s (e.g., \citealt{Ge+21}), the stereotype of S0s with poor gas is being challenged. In some SFS0s cases, the spatial distributions of molecular gas have also been resolved by the NOrthern Extended Millimeter Array (NOEMA, \citealt{Ge+20,Chen+21,Ge+21,Lu+22}). They presented that the molecular gas of SFS0s mainly distributes in the galaxy center with multiple star-forming cores, which might be triggered by gas-rich minor merger (\citealt{Ge+20, Chen+21}) or gas inflow by a stellar bar (\citealt{Lu+22}).

In the population of SFS0s,  $\lesssim$20 \% of them show lower stellar mass ($\rm M_*<10^9 M_\odot$) than normal SFS0s, namely dwarf SFS0s. \cite{Ge+21} detected the molecular gas in four dwarf SFS0s ($\rm M_{H_2}$ $\sim$$10^{6.6} M_\odot$) by Institut de Radioastronomie Millim{\'e}trique (IRAM) 30-m millimeter telescope, which was much lower than the atomic gas ($\rm M_{HI} \sim 10^{8.3} M_\odot$) detected in the literature (\citealt{Courtois+15, Haynes+18}). \cite{Ge+21} found that the kinematics of those dwarf SFS0s with faster rotation are different from normal S0s. 
Compared to certain SFS0 cases, as reported by \cite{Ge+20}, \cite{Chen+21}, and \cite{Lu+22}, dwarf SFS0s exhibit a greater variety and complexity in their multi-core structures, with star formation occurring from the center to the outskirts.
A prevalent explanation is that a series of mergers and/or accretion events in a long dynamical time scale gradually disturb the morphologies and extend the star formation history of less massive SFS0 (\citealt{Ge+21}).

\begin{figure*}
\includegraphics[width=\columnwidth]{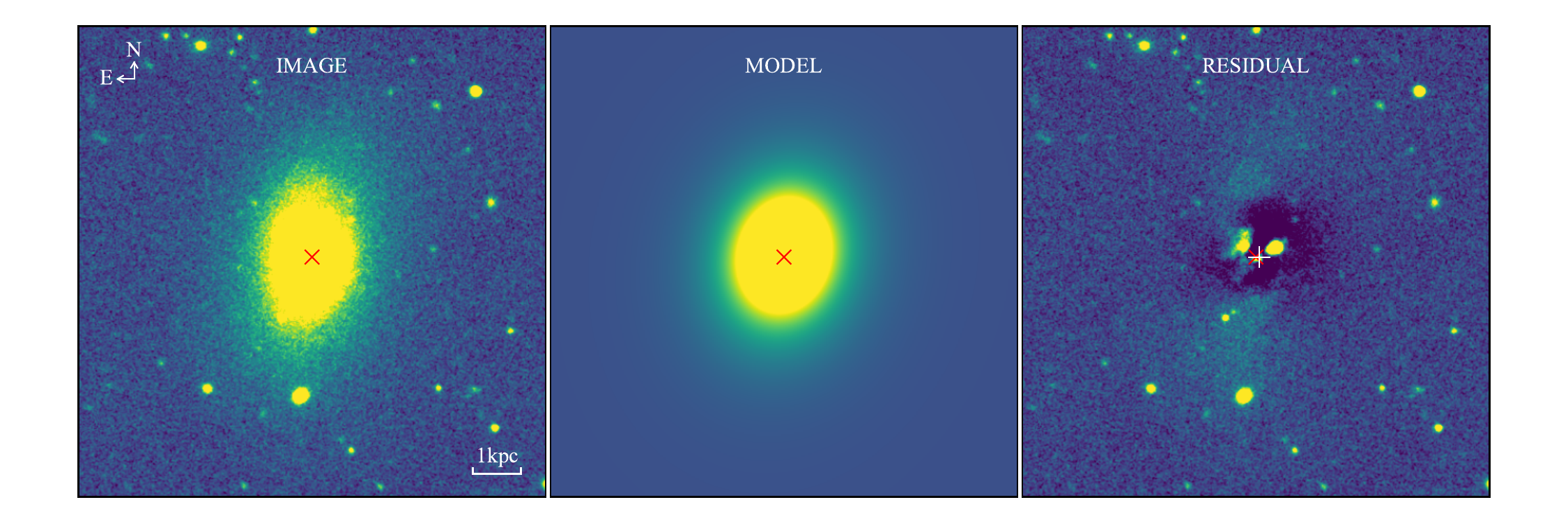}
\caption{
{\it GALFIT}-fitting results in the R band. The left to right panels present the original R-band image observed at the Palomar Observatory 60-inch telescope (\citealt{Gil+de+Paz+03}), the best-fit model by adopting a S{\'e}rsic profile, and the residual after subtracting the best-fit model from the original image, respectively. The center in the best fit is shown in the red cross, which is aligned with the galaxy center in the white plus, derived from the Ks-band isophote in Figure~\ref{fig02_image}.   
}
\label{fig01_galfit}
\end{figure*}

\begin{figure}
\includegraphics[width=\columnwidth]{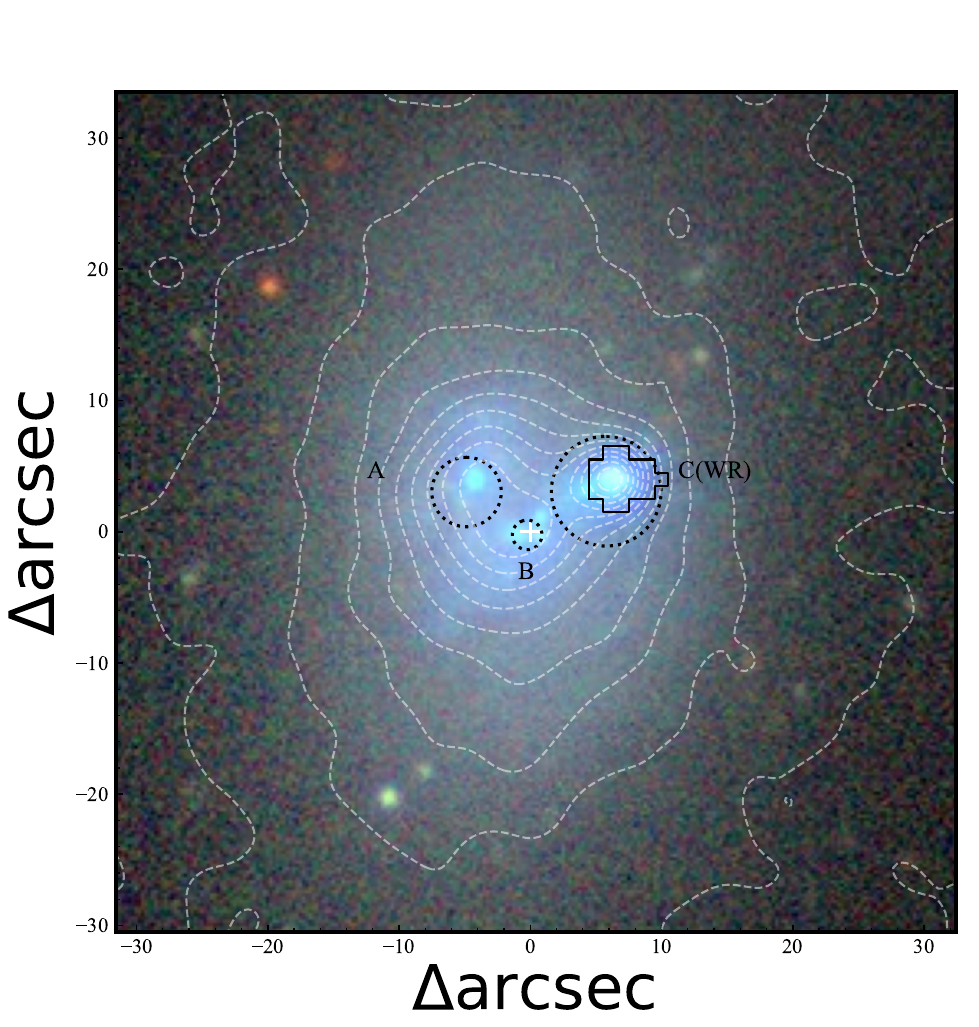}
\caption{Optical $grz$ image of PGC 44685 with triple star-forming regions from DESI Legacy Imaging Surveys (\citealt{Dey+19}). White contours show the corresponding Ks-band image observed by VISTA/VIRCAM. The closed black polygonal region represents the strongest star-forming region with Wolf-Rayet features (see section~\ref{sec3.1}). The white plus marks the center of the galaxy. Three dotted circles present three star-forming regions (labelled with A, B, and C, respectively), identified by H$\alpha$ distribution in section~\ref{sec3.3}. }
\label{fig02_image}
\end{figure}

In a sparse environment, how does a dwarf lenticular galaxy remain the active star formation, and how does the cool gas distribute?  In this work, we will study one galaxy, namely PGC 44685 (SDSS J125958.13+020257.2, or UM533) at the redshift $z$=0.00296 (i.e., the luminosity distance $\rm D_L$=12.7 Mpc).  It is classified as a lenticular galaxy by visualization or model-fitting (\citealt{de+Vaucouleurs+1991, Xiao+16, Ge+21}), and by the comparison to the Revised Hubble numerical type index T (\citealt{Naim+1995}).  Although \cite{Gil+de+Paz+03} identified it as a blue compact dwarf galaxy, they clarified that it has an outer diffuse halo with elliptical isophotes and inner irregular isophotes because of the presence of multiple star-forming regions, which is similar to the presence of star-forming lenticular galaxies in the previous studies (e.g., \citealt{Ge+20, Chen+21, Lu+22}). Therefore, \cite{Ge+21} still regarded it as the star-forming lenticular galaxy with low stellar mass, namely a dwarf SFS0 (hereafter, dSFS0).  

The left panel of Figure~\ref{fig01_galfit} shows the R-band image of PGC 44685 from \cite{Gil+de+Paz+03}, which can be used to track the stellar component. We adopted a S{\'e}rsic profile in {\it GALFIT} (\citealt{Peng+02}) to fit the R-band surface brightness by equation,
 \begin{equation}\label{eq0}
  I(r) = I_{\rm e} \exp{ \{- b_n [(r/r_{\rm e} )^{1/n} -1]\}},
 \end{equation}
 where $r_{\rm e}$ is the effective radius, $I_{\rm e}$ is the surface brightness at $r_{\rm e}$, S{\'e}rsic index, $n$, is the power-law index, which is coupled to $b_n$ ($\sim 2n-0.331$). The best-fit model and residual are shown in the middle and right panels of Figure~\ref{fig01_galfit}, and the fitting results are listed in Table~\ref{tab1}.  The S{\'e}rsic index of $n$=1.25$\pm$0.01 with an axial ratio $b/a \sim 0.8$ indicates a typical face-on disk component with symmetrical systems. The center of this symmetrical system in the R band is depicted by the red cross. 
In Figure~\ref{fig02_image},  the white contours illustrate the Ks-band isophotal distribution. The white plus is the center of this isophotal distribution.  
Both centers in the Ks band and R band are aligned with each other,  as plotted in white plus and red cross on the residual of Figure~\ref{fig01_galfit}.
Therefore, we assume the position in white plus at label B is the galaxy center and regard the galaxy as a dSFS0 in the following analysis.  In this study, we present spatially resolved optical observation with the Hispanic Astronomical Center in Andalusia (CAHA) 3.5-m telescope with the PPAK Integral Field Spectroscopy (IFS) instrument and NOEMA sub-millimeter observation for PGC 44685. As shown in the optical composite image of Figure~\ref{fig02_image}, PGC 44685 contains triple bright cores in the central region, also indicated by the residuals of Figure~\ref{fig01_galfit}. In the northwest, we find a strong star formation with the Wolf-Rayet (WR) features (shown in the black polygonal region of Figure~\ref{fig02_image}, see section~\ref{sec3.1}). We also identify triple cores as star-forming regions A, B, and C, respectively (circled by dashed lines, see section~\ref{sec3.3}).  The star formation in different regions and the total galaxy are explored for glimpsing into the triggering mechanism of star formation and the formation and evolution of low-mass lenticulars. In addition, we also study physical properties relative to the WR population.

This work is laid out as follows. Section~\ref{sec2-obs} presents CAHA optical and NOEMA sub-millimeter observations.
Section~\ref{sec3-results} shows results of the stacked WR spectrum (section~\ref{sec3.1}), the CAHA full spectral fitting (section~\ref{sec3.2}), triple star-forming regions (section~\ref{sec3.3}),  and the NOEMA CO(1-0) molecular gas (section~\ref{sec3.4}). 
Section~\ref{sec4-discussions} shows the discussions of the WR spectral constraints on the slope of initial mass function (IMF) and metallicity (section~\ref{sec4.1}), the star-forming main sequence (SFMS) and mass-metalicity relation (MZR) in different regions (section~\ref{sec4.2}), the distributions of different tracers (H$\alpha$, CO(1-0), and 3mm continuum, section~\ref{sec4.3}) and possible gas accretion scenario (section~\ref{sec4.4}). Finally, the summary is given in Section~\ref{sec5-summary}. Throughout this work, we adopt a concordance cosmology $\rm [\Omega_{\Lambda}, \Omega_{M}, h] = [0.7, 0.3, 0.7]$ and a \cite{Chabrier+03} IMF.

\begin{table}
\setlength{\tabcolsep}{1pt}
\caption{PGC 44685 physical properties}\label{tab1}
\begin{threeparttable}
\begin{tabular}{lcc}
 \hline\hline
 Property & This work     &  Literature  \\
 \hline
 R.A. (J2000)                             & 12:59:58.117 & --\\
 Del. (J2000)                              & +02:02:57.29  & --\\
 Redshift                                   & 0.00296 & 8.76 Mpc$^a$\\
R$_{\rm mag}$      & 14.11$\pm$0.05 & 13.64$\pm$0.11$^b$\\
$r_{\rm e}$ (kpc)   & 0.43$\pm$0.01&-- \\
S{\'e}rsic index $n$ & 1.25$\pm0.01$ & --\\ 
b/a  &0.80$\pm$0.01 & 0.8$^a$\\
PA (deg) & -13.91$\pm$0.48 & -- \\
 $\rm \log M_*(M_\odot)$          &  7.81$\pm$0.004   &  8.23$^a$   \\
  $\rm \log SFR (M_\odot/yr)$    & -1.40$\pm$0.002  & -1.42$^a$    \\
 $\rm \log \Sigma_{\rm SFR} (M_\odot/yr/kpc^2)$ & -1.32$\pm$0.002 & -1.42$^a$  \\
 $\rm 12+log (O/H)$                   & 8.33$\pm$0.16   &  8.37$^a$    \\
 $\rm \log L_{CO(1-0)}^{'} (K\; km\; s^{-1} pc^2)$ & 5.40$\pm$0.10 & 5.94$^a$  \\
 $\rm \log M_{\rm H_2} (M_\odot)$ & 6.04$\pm$0.16 & 6.57$^a$ \\
$\rm \log \Sigma_{M_{H_2} } (M_\odot \; pc^{-2})$ & 0.73$\pm$0.16 & --\\
 \hline
\end{tabular}
\begin{tablenotes}
 \item The superscript $^a$ indicates that all parameters are adopted from \cite{Ge+21}, where some values are converted from the \cite{Salpeter+1955} IMF to the \cite{Chabrier+03} IMF by dividing by a factor of 1.7. The superscript $^b$ indicates that the R-band magnitude is from \cite{Gil+de+Paz+03}. PA is the position angle by increasing counter-clockwise (from North to East).
 \end{tablenotes}
\end{threeparttable}
\end{table}

\section{Observations}
\label{sec2-obs}

\subsection{CAHA }
\label{sec2.1-caha} 
The optical IFU spectroscopic observation of PGC 44685 was performed with the Potsdam Multi Aperture Spectrograph (PMAS), mounted on the CAHA 3.5-m telescope at the Calar Alto observatory. The PPAK (PMAS fiber Package) fiber bundle offers a large (74$\arcsec \times$64$\arcsec$) hexagonal field of view (FoV), which covers the whole galaxy up to 2-3 times effective radii (\citealt{Verheijen+04, Kelz+06}). PGC 44685 was observed on 2017 April 2, with the low-resolution (V500; R$\sim$850) setup covering the wavelength range of 3745–7500$\rm \AA$. A three-pointing dithering scheme was taken to obtain a filling factor of 100\% across the entire FoV. The exposure time per pointing was 900 seconds at a sampling of 1$\rm \arcsec \times$1$\arcsec$ per spaxel. 

The PPAK IFU data were reduced by an upgraded Python-based pipeline (\citealt{Garcia-Benito+15, Sachez+16}). An outline reduction process includes several steps, such as identifying the spectral location on the detector, extracting individual spectra and correcting their distortions, calibrating the wavelength, correcting fiber transmission and flux, subtracting the sky, and reconstructing a final 3D datacube. We recommend readers to the literature for more detailed data reduction (\citealt{Husemann+13, Garcia-Benito+15}). Note that the resampled spaxel size is smaller than the effective spatial resolution of the PPAK datacube.

\subsection{NOEMA}
\label{sec2.2-noema}
The CO(1-0) molecular gas was observed three times with NOEMA on 2021 April 15, May 05, and May 13, respectively (project W20BV; PI: Shiying LU). It was observed in the D configuration with 10 antennas in dual polarizations, covering the spectral window from 93.4 to 116.0 GHz. The total on-source time was 9.3 hours. The object, 3C273, was used as the receiver bandpass (RF) and the phase/amplitude calibrators simultaneously. 

Data reduction was calibrated by Continuum and Line Interferometer Calibration (CLIC), a module of the Grenoble Image and Line Data Analysis Software (GILDAS), including RF, flux, phase, and amplitude calibrations. The other module in GIDLAS, MAPPING, was adopted to clean and image the observed data, including the 3mm dust continuum and CO(1-0) emission line in the $uv$ plane. For the CO(1-0) emission line, the redshifted frequency is 114.931 GHz, given its redshift of $z=$0.00296 ($\nu_{\rm rest}=$115.271 GHz). We output a CO(1-0) datacube, which covers a velocity range from -506.07 to 537.37 km/s at a velocity resolution of 5.2 km/s ($\sim$2 MHz, the coarsest spectral resolution of NOEMA).  The synthesized beam size is 4.20$\arcsec \times$3.14$\arcsec$ with a position angle (PA) of -2.06$^{\circ}$. The CO(1-0) datacube has a map cell of 0.56$\arcsec \times$0.56$\arcsec$ and covers an FoV of  43.9$\arcsec \times$43.9$\arcsec$ with a noise ($\rm \sigma_{rms}$) of 1.85 mJy/beam. For the rest 3mm dust continuum behind the emission line, we output the integrated image covering an FoV of 53.2$\arcsec \times$53.2$\arcsec$ with a noise of 9.40$\mu$Jy/beam. The corresponding beam size is 4.84$\arcsec \times$3.61$\arcsec$ at a PA of -1.34$^{\circ}$. The continuum map cell is the same as the CO(1-0) emission line.

\section{Data Analysis and Results} \label{sec3-results}
\subsection{Wolf-Rayet (WR) Region} \label{sec3.1} 

\begin{figure*}
\includegraphics[width=0.9\textwidth, angle=0 ]{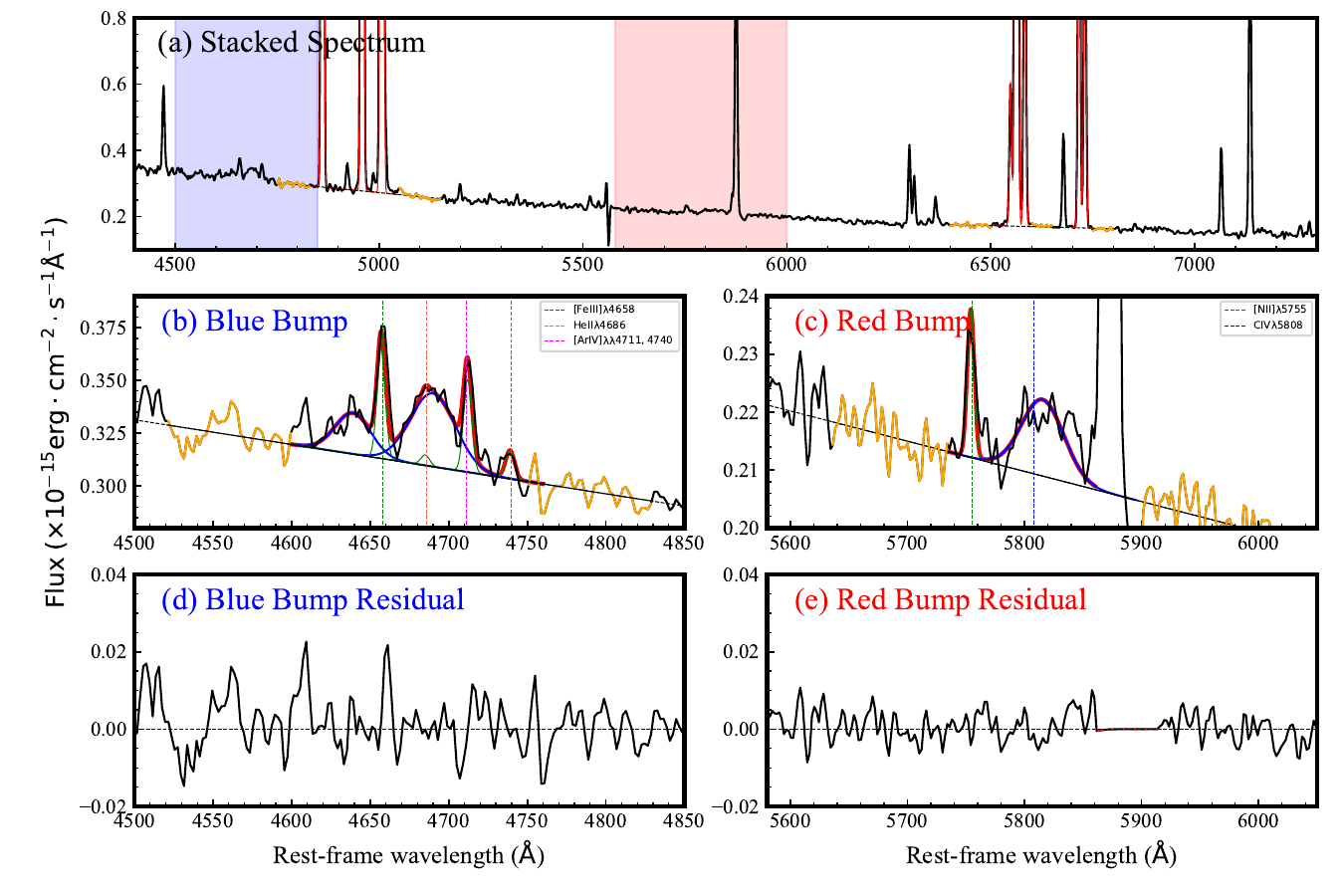}
\caption{The stacked WR spectrum (in a black curve) summing all 20 spectra with WR features. Panel (a) shows the spectrum in a large range of wavelengths. The shaded blue and red regions represent WR blue and red bump regions, respectively. The zoom-in spectra are shown in panels (b) and (c), separately. Orange curves are windows used to fit the continuum. The best linear fitting indicates the best-fit continuum, shown in the black line. Broad Gaussian components in solid blue curves and narrow Gaussian components in solid green curves are overlaid on the best-fitting continuum. The red lines imply the final best-fit profiles, i.e., all Gaussian components plus the linear-fit continuum. Colored vertical dashed lines illustrate the wavelength centers of emission lines within the bump ranges, and corresponding legends are printed at the top-right of each panel. Panels (d) and (e) show their residual spectra after subtracting the continua and Gaussian components. The clear residuals imply that the fitting is appropriate.}
\label{fig03_WR}
\end{figure*}

Thanks to the optical IFU data from CAHA, we can obtain the spectrum for each spatial spaxel within the FoV of a galaxy, which reveals more information about the corresponding position. Active star formation typically coincides with strong ionized emission lines, such as the $\rm H\alpha, H\beta$, [OIII]$\lambda\lambda4959,5007$, [NII]$\lambda\lambda6548,6583$, and [SII]$\lambda \lambda 6717, 6731$ lines. These lines are commonly used to trace the attenuation, determine gas-phase metallicity, and classify star formation activity. While the direct signatures of WR stars, shown in a broad HeII$\lambda4686$ (blue bump) and/or CIV$\lambda5808$ (red bump) emission features produced by their dense stellar winds (\citealt{Crowther+07}), are relatively rare due to their small numbers at birth and short lifetime. WR stars are believed to be the evolved descendants of massive stars with initial mass $\rm M_*>25 M_\odot$ for solar metallicity and a lifetime of 2-5 Myr. The existence of the region with WR-star signatures (i.e., WR regions) can clearly indicate that the host galaxy is experiencing ongoing or very recent violent star formation, which theoretically accompanies strong stellar winds. In PGC 44685, we clearly detect the WR emission lines, and introduce the stacked WR spectrum and the number of WR stars in sections \ref{sec3.1.2} and \ref{sec3.1.3}, respectively.

\subsubsection{Identification of WR region} \label{sec3.1.1}
In the optical spectrum, the most prominent emission feature in WR regions is the blue bump covering 4600-4750 \AA\ (\citealt{Schaerer+Vacca+1998}), which is a complex emission combination formed from the blend of broad stellar lines, such as HeII$\lambda4686$ (dominated), NIII$\lambda \lambda \lambda 4628,4634,4640$ and  CIII/CIV$\lambda \lambda4650,4658$(\citealt{Guseva+2000, Crowther+07}),  and a number of narrow nebular emission lines, like HeII$\lambda 4686$, [FeIII]$\lambda \lambda \lambda 4658,4665,4703$ and [ArIV]$\lambda\lambda4711,4740$ (\citealt{Guseva+2000}). Besides the WR blue bump, there is the other broad emission feature, called the red bump centered at 5808\AA, formed mainly by the broad CIV$\lambda \lambda5801, 5812$ stellar lines, and superposed by narrow auroral emission line [NII]$\lambda5755$. Compared to H$\beta$, the WR blue bump is generally faint, only a few percent of H$\beta$ flux, and the red bump is usually even fainter than the blue bump. In order to search for WR features of each spaxel, we select the blue bump window as the proxy of the tracer and adopt the method mentioned in \cite{Miralles-Caballero+16} (hereafter, MC16) to search for all spaxels and to pick out all spectra with WR features. 
 
The detailed process of WR spectral selection is mainly divided into three steps. We specify a portion of the spectrum spanning the rest-frame range of 4500-4850$\rm \AA$ to search for all resolved spectra with WR features, particularly covering the blue bump range. For each spaxel, the first step is to select a spectral window on each side of the blue bump. One is the wavelength covering 4520-4600\AA, and the other is 4750-4830\AA. The corresponding windows are also shown as orange lines in panel (b) of Figure~\ref{fig03_WR}. We employ a linear fit to obtain the continuum and then subtract it from the original spectrum. The second step is to calculate the detection significance, $\varepsilon$, of the blue bump in each spectrum. We integrate the flux density, $F_{\rm bump}$, within 4630-4730\AA, and compare it with the root mean square in two spectral windows. Following \cite{Tresse+1999}, the definition of the detection significance, $\varepsilon$, is as 
\begin{equation}\label{eq1}
 \varepsilon = \frac{F_{\rm bump}}{\sigma}=\frac{F_{\rm bump}}{\sigma_c D \sqrt{2N+\frac{EW}{D}}} \approx \frac{F_{\rm bump}}{\sigma_c D \sqrt{2N}} ,
\end{equation}
where $\sigma_c$ is the standard deviation for two sides of the windows continuum, D$\rm =2\ \AA$/pixel is the reciprocal dispersion, $N$ is the number of spectral points in the integration of flux $F_{\rm bump}$, and $EW$ is the equivalent width of the blue bump. Considering that the $EW$ contribution in this function is low, we can adopt an approximation. In the third step, we adopt the significance $\varepsilon>5$ as the selection criterion to select 23 spaxels. After excluding 3 spaxles located outside the WR region, we finally pick out 20 spaxel's spectra, which mainly distribute on the brightest core in the northwest, shown in a black polygon of Figure~\ref{fig02_image}.
 
\subsubsection{Stacked WR spectrum} \label{sec3.1.2}
Considering the weak blue bump in each spaxel, we sum 20 spectra with WR features in the rest frame to obtain a stacked WR spectrum with a high signal-to-noise ratio (SNR), shown in panel (a) of Figure~\ref{fig03_WR}. The blue and red bump ranges are highlighted in light-shaded blue and red regions, respectively. Other emission lines, such as H$\alpha$, H$\beta$, [OIII]$\lambda \lambda 4959,5007$, [NII]$\lambda \lambda 6548,6583$ and [SII]$\lambda\lambda6717,6731$, are comparably stronger than WR blue/red bumps. Since most single stellar population (SSP) templates, such as Medium resolution INT Library of Empirical Spectra (MILES, \citealt{Sanchez-Blazquez+06, Vazdekis+10}) do not consider the WR phase, inappropriate subtraction of continuum adopting those templates could weaken the features of the blue and even fainter red bumps. Therefore, in panel (a) of Figure~\ref{fig03_WR}, we adopt a linear function (in a black straight line) within spectral windows (in orange curves) to represent the part of the stacked continuum. 
Superimposed on this linear fitting, we adopt a single Gaussian profile to fit each strong emission line. 
The H$\beta$ is fitted together with the [OIII]$\lambda \lambda 4959,5007$ line, H$\alpha$ is fitted together with the [NII]$\lambda \lambda 6548,6583$ line, and the [SII]$\lambda\lambda6717,6731$ double lines are fitted together.

Employing the best Gaussian fit in the WR region, depicted by red curves in panel (a) of Figure~\ref{fig03_WR}, we calculate the H$\alpha$ to H$\beta$ ratio to derive the nebular extinction. This enables us to estimate the star formation rate (SFR) using the extinction-corrected  H$\alpha$ luminosity, following \cite{Kennicutt+1998}:
\begin{equation}\label{eq2}
 {\rm SFR}[M_\odot {\rm yr^{-1}}] = 7.9 \times 10^{-42} L({\rm H}\alpha).
\end{equation}
Note that SFR was derived based on the \cite{Salpeter+1955} IMF, we thus convert it based on \cite{Chabrier+03} by dividing a factor of 1.7.
Then, the inclination-corrected surface density of SFR ($\rm \Sigma_{SFR}$) could be obtained by considering the axial ratio of $q_0=0.25$ for S0 galaxies (\citealt{Sandage+1970}). Even when considering $0.2 < q_0 < 0.5$, the impact on the estimation of $\Sigma_{\text{SFR}}$ is less than 10\%, which does not affect the main results of this work.
To estimate the number of each type of WR stars, the metallicity in the WR region is needed in section~\ref{sec3.1.3}. To ascertain the consistency of the method mentioned in \cite{Miralles-Caballero+16}, we also adopt strong line calibrations based on easily observable, optical lines for the direct determination of metallicity. But instead of the O3N2 method used in \cite{Miralles-Caballero+16}, we prefer to adopt the RS32 method in \cite{Curti+20} to estimate the gas-phase metallicity, because of its weakly dependent on the ionization parameter. The gas-phase metallicity, 12+log(O/H), based on the RS32 method can be obtained by the formula given by \cite{Curti+20}:
\begin{equation}\label{eq3}
\rm \log (RS32) = \Sigma_{N} c_n [12+\log(O/H) -8.69 ]^{n},
\end{equation}
where RS32$=\rm \frac{[OIII]\lambda5007}{H\beta} +  \frac{[SII]\lambda\lambda6717,6731}{H\alpha}$. The $c_{\rm n}$ parameter is taken from \cite{Curti+20}. All parameters in the WR region are presented in Table~\ref{tab2}. Within the validation of metallicity based on the O3N2 method (i.e., $\rm 12+\log(O/H)>$8.1), we also check the metallicity using the same method in \cite{Miralles-Caballero+16}, which is broadly ($<0.4$ dex) consistent with the value based on the RS32 method considering the method differences. The metallicity using different methods does not significantly affect the number estimation of WR types in section~\ref{sec3.1.3}. Since we have already proposed the GTC/MEGARA observation in the WR region (also mentioned in section~\ref{sec4.3}), a more detailed analysis of the gas will be available in our next paper based on the new higher-resolution dataset.

Given the current spectral quality, it is impossible to fit so many features in blue and red bumps independently and accurately. 
We refer to some thoughts and follow methods in the literature (\citealt{Lopez-Sanchez+10, Miralles-Caballero+16, Liang+20, Liang+21}) to consider the occurrence of each feature and its relative strength. We determine that for the blue bump in panel (b) of Figure~\ref{fig03_WR}, two broad Gaussian profiles (in blue lines) plus four narrow ones (in green lines) are finally employed to fit the broad blend of carbon and nitrogen components at the centroid of 4650\AA, (hereafter, called the broad C/N$\lambda$4650 mixture), the broad HeII$\lambda4686$ component, and the narrow nebular [FeIII]$\lambda4658$, HeII$\lambda 4686$, and [ArIV]$\lambda \lambda4711,4740$ emission lines, respectively. For the red bump in panel (c) of Figure~\ref{fig03_WR}, we firstly mask a stronger neighbouring narrow HeI$\lambda5875$ for better appropriate fitting, then employ one broad plus one narrow Gaussian profile to fit the broad CIV$\lambda5808$ component and narrow [NII]$\lambda5755$ line, separately. For all broad components in the blue and red bumps, the Gaussian centers are allowed to vary within $\pm$7.5\AA, and the Gaussian widths $\sigma$ are free parameters within the range of 3-22\AA, where the upper limit corresponds to the velocity dispersion of 1425 km/s, the largest possible value for individual WR-stars winds (\citealt{Willis+04}). For all narrow nebular lines, the Gaussian widths $\sigma$ are tied to that ($\sim$3\AA) of H$\beta$, and the centers are fixed at the known wavelength. The best-fitting results for blue and red bumps are shown in red curves in panels (b) and (c) of Figure~\ref{fig03_WR}. The bottom panels (d) and (e) present their corresponding residual spectra. Finally, we estimate the velocity of the broad HeII$\lambda4686$ blue bump, $v_{\rm BB}\sim+320$km/s with $\sigma=$15\AA, and the velocity of the broad CIV$\lambda5808$ red bump, $v_{\rm RB}\sim +410$km/s with $\sigma=$18\AA, which indicate that the WR stellar winds are not so strong. 
 
\subsubsection{Number of WR-stars} \label{sec3.1.3}
Based on the best fitting of the stacked WR spectrum, the physical properties (i.e., flux, velocity, and velocity dispersion) of each broad component have been obtained, which can be used to calculate the number of WR stars. In the blue bump range, the broad HeII$\lambda4686$ bump and emission lines are mainly linked to the nitrogen-rich WR (i.e., WN) stars, while the contribution from carbon-rich WR (i.e., WC) stars is also included in the blue bump. Thus, the blue bump is attributed to all WR types. In the red bump range, the broad CIV$\lambda5808$ bump essentially is contributed by the WC stars, which are in the early type of the WR phase and are not easy to detect due to their weakness. No matter whether WN or WC stars, both can contribute to the broad C/N$\lambda$4650 mixture. Thus, it's not easy to separate and calculate the number of each type of WR star exactly. 
Given that we can not obviously detect both the NV$\lambda \lambda4605, 4620$ emissions from WN early-type stars and the CIII$\lambda5696$ emission mainly from the WC late-type stars, we assume that emissions in bumps are contributed from two types of WR stars,  i.e., the WN late-type (i.e., WNL) and the WC early-type (i.e., WCE) stars. The WNL stars mainly contribute to the broad HeII$\lambda 4686$ component, while the WCE stars prominently dominate the broad CIV$\lambda5808$ bump. 
Although the lifetime of WR stars is short, their detections with IFUs have been reported in the literature (\citealt{Garcia-Benito+10, Liang+20}). To calculate the numbers of WNL and WCE stars for PGC 44685 in the WR region, we mainly refer to the literature of \cite{Miralles-Caballero+16}, which gives a rough estimation of WCE's number by solving two equations even if the red bump is weak or non-detected. We also employ the other method from \cite{Lopez-Sanchez+10} (hereafter, LS10) to speculate the number of WCEs directly.

\begin{table}
\centering
\caption{Properties of the WR region in PGC 44685.}\label{tab2}
\begin{threeparttable}
\begin{tabular}{lcr}
 \hline\hline
 Property & Unit/Method & Value \\
 \hline
 $\log M_{*}^{\rm WR}$ & $\rm M_\odot$ & 6.67$\pm$0.014 \\
 $\log \rm SFR$ & $\rm M_\odot /yr$ & -1.66$\pm$0.002 \\
 $\log \Sigma_{\rm SFR}$ & $\rm M_\odot /yr/kpc^2$ & -0.42$\pm$0.002 \\
 H$\alpha$/H$\beta$ & -- & 2.86$\pm$ 0.05 \\
$\rm 12+log(O/H)$& -- & 7.71$\pm$ 0.16\\
$\rm \log M_{H_2} $ & $\rm M_\odot $ & 5.74$\pm$0.14 \\
$\rm \log \Sigma_{M_{H_2} } $ & $\rm M_\odot \; pc^{-2}$ & 0.75$\pm$0.14 \\
EW(H$\beta$) &\AA & 163.74$\pm$0.43 \\
 $L(4650)^{\rm a}$ & erg/s &(1.05$\pm$ 0.43)$\times 10^{37}$ \\
 $L(4686)^{\rm b}$ & erg/s &(2.37$\pm$0.81)$\times 10^{37}$ \\
 $L(5808)^{\rm c}$ & erg/s &(1.63$\pm$4.19)$\times 10^{36}$  \\
 $N_{\rm WNL}$ & MC16 & 26$\pm$11 \\
                      & LS10  & 29 $\pm$11\\
 $N_{\rm WCE}$ & MC16 & 3$\pm$ 1\\
                        &LS10 & 1$\pm$3\\
  $N_{\rm O\text -star}$ & MC16 & 270$\pm$30 \\           
 \hline
\end{tabular}
 \begin{tablenotes}
 \item The superscripts $\rm ^{a,b,c}$ denote the luminosities of three broad Gaussian components (shown in blue curves in Figure~\ref{fig03_WR}), i.e, the broad C/N mixture, the broad HeII$\lambda$4686 blue bump, and the broad CIV$\lambda5808$ red bump, respectively.
 \end{tablenotes}
\end{threeparttable}
\end{table}

\begin{figure*}
\includegraphics[width=0.9\textwidth, angle=0 ]{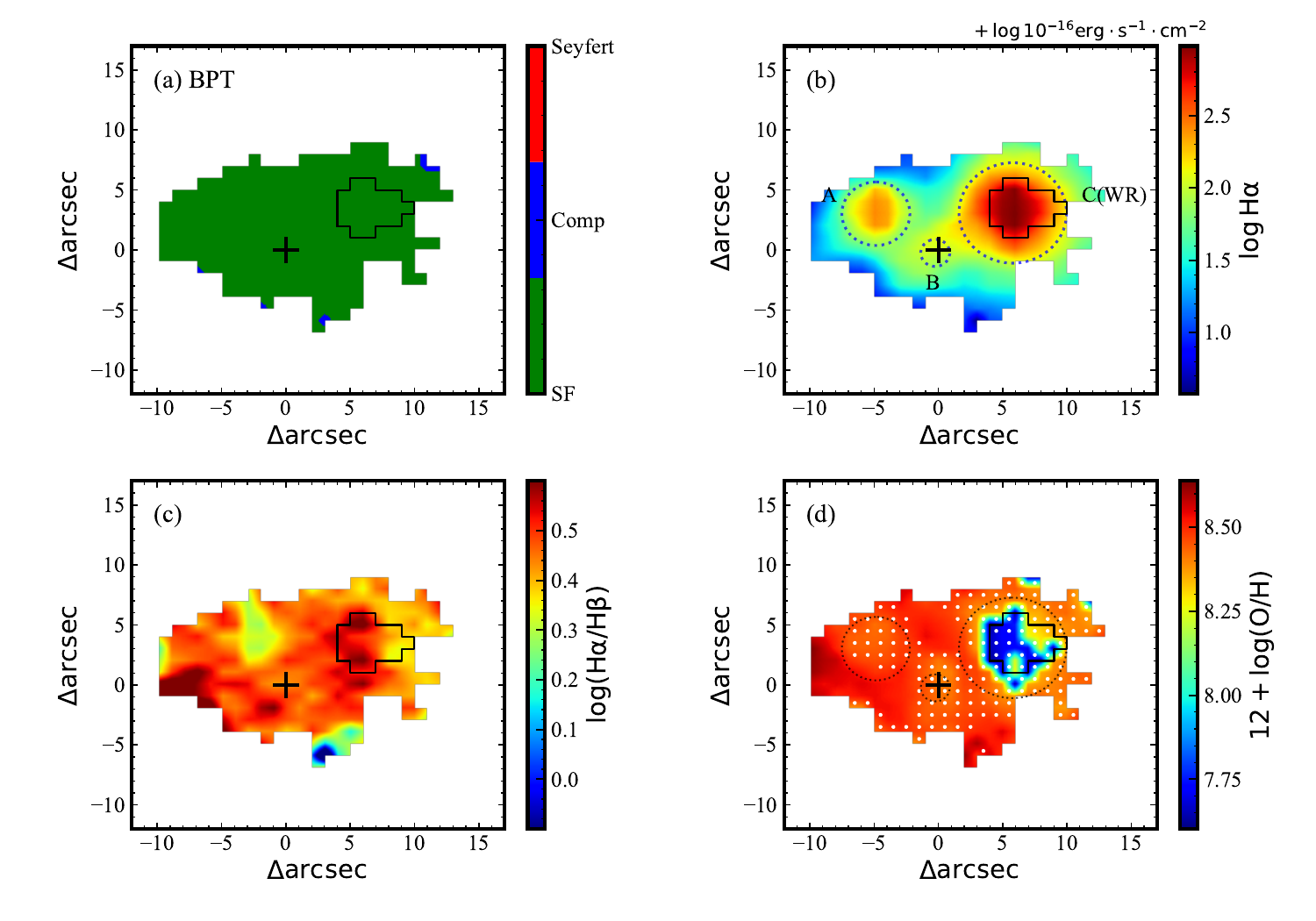}
\caption{Distributions of the physical parameters color-coded by the BPT diagram (panel a), H$\alpha$ flux (panel b), the ratio of H$\alpha$ to H$\beta$ (panel c), and gas-phase metallicity based on the RS32 method (panel d), respectively. All spaxels satisfy the criteria of SNR$>5$ for both stellar and emission components. The black plus marks the galaxy center. The black polygons mark the found WR region (see section~\ref{sec3.1}). Three dotted circles in panels (b) and (d) present triple star-forming regions with labels A, B, and C, identified by H$\alpha$ distribution in section~\ref{sec3.3}. The WR region is located within the region C. White dots in panel (d) illustrate the anomalously low-metallicity (ALM) region with $\rm \Delta \log(O/H)<-0.185$ (see section~\ref{sec3.2} and Figure~\ref{fig05_ALM} for details).}  
\label{fig04_opt}
\end{figure*}

Based on the measured fluxes of the broad C/N$\lambda4650$ mixture and the broad HeII$\lambda$4686 components, we can calculate the corresponding luminosities, i.e., $L$(4650)$=(1.05\pm0.43) \times 10^{37}$ erg/s, and $L$(4686)$=(2.37\pm0.81)\times 10^{37}$ erg/s. To drive the  numbers of WNL  and WCE stars, corresponding to the two unknowns $x$ and $y$, respectively, we can solve a system of two equations in MC16 method as follows: 
\begin{eqnarray} \label{eq4}
\Biggl\{
\begin{array}{ll}
 L(4686)=x \times L_{\rm WNL}({\rm HeII}) + y \times L_{\rm WCE}({\rm HeII}),  \\
L(4650)=x \times L_{\rm WNL}({\rm NIII})  + y \times L_{\rm WCE}({\rm CIII/CIV}),
\end{array}
\end{eqnarray}
where $L$(4686) considers the luminosity of a single WNL star, $L_{\rm WNL}$(HeII), and the luminosity of a single WCE star, $L_{\rm WCE}$(HeII) in the broad HeII$\lambda$4686 range; $L$(4650) sums the luminosities from a single WNL star ($L_{\rm WNL}$(NIII)$\rm =3.8\times 10^{35}$ erg/s, \citealt{Crowther+Hadfield+06}) and from a single WCE star, $L_{\rm WCE}$(CIII/CIV), within the range of the broad C/N mixture. The $L_{\rm WNL}$(HeII) can be estimated by a function of metallicity, proposed in the LS10 approach, 
\begin{equation} \label{eq5}
L_{\rm WNL}({\rm HeII}) = (-5.430 + 0.812 \times Z_{\rm gas}) \times 10^{36} \rm erg/s,
\end{equation}
where $\rm Z_{gas}= 12+log(O/H)$ is the gas-phase metallicity, listed in Table~\ref{tab2}. Within the WR region (the black polygon in Figure~\ref{fig02_image}), we can calculate the $L_{\rm WNL}(\rm HeII)=(0.83\pm 0.13)\times 10^{36}$ erg/s. Following the MC16 approach, we assign 12\% of the luminosity of LMC WC4 stars ($4.9 \times 10^{36}$ erg/s, reported in \citealt{Crowther+Hadfield+06}) to the WCE contribution in the broad HeII$\lambda$4686, i.e, $L_{\rm WCE}(\rm HeII)=(0.12\times4.9) \times 10^{36} $ reg/s, whereas the remaining 88\% is for the $L_{\rm WCE}(\rm CIII/CIV$).  Finally, the derived number of WNL stars is $N_{\rm WNL}=26\pm11$, and the number of WCE stars is $N_{\rm WCE}=3\pm1$, listed in Table~\ref{tab2}.

Panel (c) of Figure~\ref{fig03_WR} implies the detection of the broad CIV$\lambda5808$ red bump, so $N_{\rm WCE}$ can be estimated by the LS10 approach again, which is equal to the ratio of the observed luminosity in the broad CIV$\lambda5808$ red bump to the $Z_{\rm gas}$-based one, 
\begin{equation} \label{eq6}
N_{\rm WCE}=\frac{L_{\rm obs}(\rm CIV\lambda5808)}{ L_{\rm WCE}(\rm CIV\lambda5808)},
\end{equation} 
with $L_{\rm WCE}(\rm CIV\lambda5808)=(-8.198 +1.235 \times Z_{\rm gas}) \times 10^{36}$ erg/s. When the broad HeII$\lambda4686$ blue bump is substituted for the broad CIV$\lambda5808$ red bump employing this similar approach plus the equation (\ref{eq5}), the $N_{\rm WNL}$ also can be assessed again. Those derived numbers adopting the LS10 method are shown in Table~\ref{tab2}. Within the error, both the LS10 and MC16 approaches give us consistent results and support the domination of WNL stars in the WR region of PGC 44685.

Following the MC16, we can simply evaluate the number of O stars ($N_{\rm O\text -star}$) by
\begin{equation} \label{eq7}
N_{\rm O\text -star}=\frac{Q_0^{\rm Total}-N_{\rm WNL}Q_0^{\rm WNL}-N_{\rm WCE}Q_0^{\rm WCE}}{\eta_0(t)Q_0^{\rm O7V}},
\end{equation}
where $Q_0^{\rm WNL}$ and $Q_0^{\rm WCE}$ denote average numbers of ionizing photons of a WNL and a WCE star, respectively. As reported by \cite{Crowther+Hadfield+06}, we assume the averages of $\log Q_0^{\rm WNL} = 49.4$ and $\log Q_0^{\rm WCE}=49.5$, which do not vary with metallicity. The $Q_0^{\rm O7V}$ number of ionizing photons in the $\log$ scale is $\log Q_0^{\rm O7V}=49.0$ reported by \cite{Schaerer+Vacca+1998}, and the total number of ionizing photons is, $Q_0^{\rm Total}=N_{O'}Q_0^{\rm O7V}$, where $N_{O'}=L({\rm H\beta})/L(\rm O7V, H\beta)$, assuming the H$\beta$ luminosity of O7V star, $L(\rm O7V, H\beta)=4.76\times 10^{36}$ erg/s. 
Here we assume the $\eta_0(t)=1.0$ and adopt the $N_{\rm WNL}$ and $N_{\rm WCE}$ in the MC16 approach. Finally, we obtained the number of O stars $<$ 300, as shown in Table~\ref{tab2}. Notice that we have already taken into account any potential impact of uncertainties on the number etimations of WNL, WCE, and O stars, such as the possible 15\% uncertainty stemming from the absolute photometric calibration based on the analysis of the Calar Alto Legacy Integral Field Area survey (\citealt{Miralles-Caballero+16}), the total intrinsic scatter ($\pm$0.15 dex) in the relation between the RS32 calibrator and the gas metallicity (\citealt{Curti+20}) and the large redshift uncertainty ($\Delta z\sim$0.00044, i.e., 2 Mpc) comparing to the highest estimation in the literature (\citealt{Fairall+1980}, z$\sim$0.0034).

\begin{figure*}
\includegraphics[width=0.8\textwidth, angle=0 ]{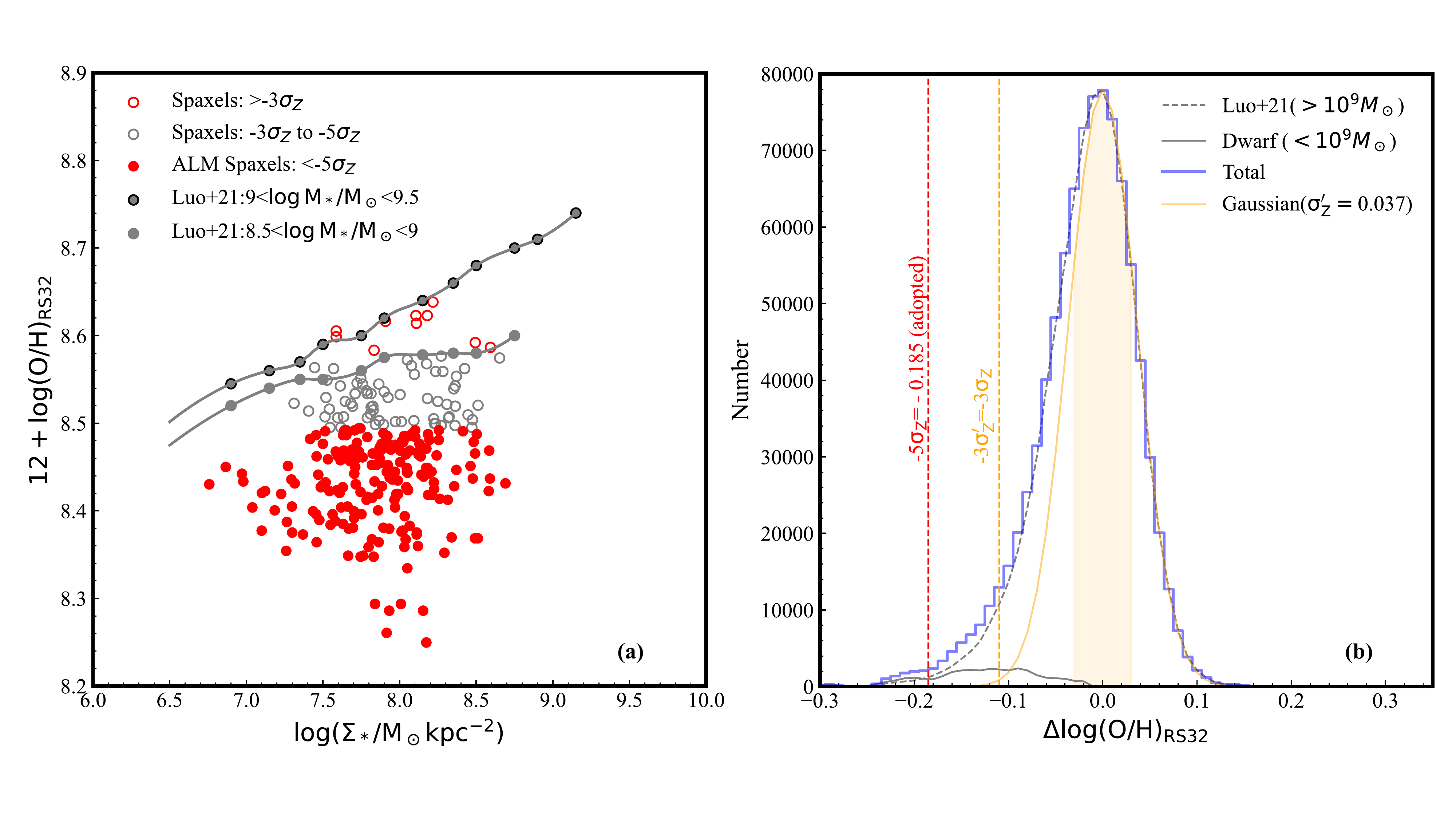}
\caption{Left panel a:  Relationship between metallicity and stellar surface mass density ($\Sigma_*$-Z$_{\rm gas}$). 
The solid red circles represent spaxels belonging to the anomalously low-metallicity (ALM) region, satisfying the criterion of $\rm \Delta \log(O/H)<-5\sigma_Z$ ($\sigma_Z=$0.037 from \citealt{Luo+21}), which is corresponding to the red vertical dashed line in the right panel b. The open red and grey circles show the spaxels with $\rm \Delta \log(O/H)> -3\sigma_Z$ and $\rm -3\sigma_Z<\Delta \log(O/H)< -5\sigma_Z$, respectively. The grey dots and curves are from \cite{Luo+21} in two low-mass bins.  
Right panel b: Distribution of the metallicity deviation $\rm \Delta \log(O/H)$ in star-forming spaxels. The grey dashed curve presents the distribution in galaxies with $M_*>10^9 M_\odot$, which is compiled from \cite{Luo+21}. 
The grey solid curve shows the distribution in dwarf ($M_* < 10^9 M_\odot$) galaxies, assuming that $\sim$2\% of dwarf galaxies in the SDSS-IV MaNGA survey behave like PGC44685. 
The blue histogram is the total distribution of both. 
We fit only the positive side of the blue histogram to derive the symmetric negative side using a Gaussian profile.
The orange line shows the best-fit Gaussian profile, with a standard deviation $\sigma^{\prime}_Z=$0.037, which is the same value ($\sigma_Z=$0.037) obtained in \cite{Luo+21}. The shaded region shows the $\pm \sigma^{\prime}_Z$  coverage around the center of Gaussian distribution.
The vertical dashed lines mark the locations where $\rm \Delta \log(O/H) = -5\sigma_Z$ (red) and where $\rm \Delta \log(O/H)=-3\sigma^{\prime}_Z=3\sigma_Z$ (orange), respectively. We refer to those spaxels that deviate from a lower metallicity by more than the former value as ALM spaxels.  }
\label{fig05_ALM}
\end{figure*}

\subsection{CAHA full spectra fitting} \label{sec3.2}

The CAHA optical observation allows us to study the spatial distribution of the physical properties of the galaxy. We adopt a public full spectral fitting approach, Penalized Pixel-Fitting (pPXF, \citealt{Cappellari+Emsellem+04, Cappellari+17}), to fit the stellar continuum after masking the emission lines. For all spaxels, spectra with SNR$>$5 are selected and then fitted with SSP templates of MILES (\citealt{Vazdekis+10}), assuming a \cite{Chabrier+03} IMF and \cite{Calzetti+2000} dust extinction. The SSP templates cover the 25 ages from 0.06 Gyr to 15.85 Gyr and six metallicities (i.e., $\rm \log [M/H]=$ -1.71, -1.31, -0.71, -0.4, 0.0, 0.22). The distorted stellar velocity distribution can be obtained, and similar results have been shown in Fig.9 of \cite{Ge+21}. After subtracting the best-fit synthesized stellar continuum to obtain the pure emission lines, we adopt a Gaussian profile to fit each emission line, similar to that mentioned in section~\ref{sec3.1.2}. Adopting the same approach in \cite{Lu+22} to estimate the flux and noise of each emission line, we finally select $\sim$300 spectra with the SNR of emission lines larger than 5, which mainly cover the central region of PGC 44685, as shown in Figure~\ref{fig04_opt}. 
Note that since common emission lines (such as H$\alpha$, H$\beta$, [OIII], and [SII]) are much stronger than WR bumps,  we verified that these line measurements even in the WR region can not be significantly affected by adopting the lack of a proper WR phase models (i.e., MILES). 
Within this coverage, the stellar mass, SFR, $\rm \Sigma_{SFR}$, and 12+log(O/H) can be assessed by the same functions mentioned in section~\ref{sec3.1.2}, and results are shown in Table~\ref{tab1}. Compared to the results of \cite{Ge+21}, our derived stellar mass and SFR are smaller due to the smaller coverage.
We also stacked the pure emission lines within the WR region to re-assess the physical properties instead of adopting the line-fitting continuum mentioned in section~\ref{sec3.1.2}. As expected, both results are consistent. 

\begin{table*}
\caption{Physical properties of stacked ALM spectra in three star-forming regions.} \label{tab3}
\begin{threeparttable}
\begin{tabular}{l|cccccccc}
 \hline\hline
 Region & $\log M_*  $ & $\log \rm SFR$  &  $\log \rm \Sigma_{SFR} $   &  $\rm 12+log(O/H)$ &  $L_{\rm CO}$ & $\log M_{\rm H_2}$  & $\rm \log \Sigma_{M_{H_2}}$ &EW(H$\beta$)$^b$ \\
 \cline{2-9}
  ($\rm N^{a}_{pixel}$)  &  $\rm M_\odot$   & $\rm M_\odot /yr$  & $\rm M_\odot /yr/kpc^2$   &   RS32-based  & $ \rm K\; km\; s^{-1} pc^2 $ & $\rm M_\odot$ & $\rm M_\odot \; pc^{-2}$& \AA   \\
 \hline
   A (20)  & 6.87$\pm$0.015 &  -2.51$\pm$0.002  & -1.27$\pm$0.002    & 8.44$\pm$0.15  & 4.67$\pm$0.14  &  5.30$\pm$0.20  & 0.38$\pm$0.19 & 41.82$\pm$0.02\\
   B (4)   & 6.01$\pm$0.021  &  -3.25$\pm$0.003  &  -1.31$\pm$0.003  &  8.40$\pm$0.16 &  4.29$\pm$0.08  &  4.92$\pm$0.15 & 0.67$\pm$0.15 &22.57$\pm$0.09\\
   C (54) & 7.12$\pm$0.009  & -1.55$\pm$0.003  &  -0.75$\pm$0.003   &  7.67$\pm$0.16  & 5.36$\pm$0.05  &  6.00$\pm$0.14 & 0.58$\pm$0.14 & 107.56$\pm$0.01\\  
 \hline
\end{tabular}
 \begin{tablenotes}
  \item  The superscript $^a$ represents the number of spaxels counted in the optical CAHA IFU data, and each spaxel belongs to the anomalously low metallicity (ALM) region. The superscript $^b$ presents the equivalent width of H$\beta$, measured within $\pm$3$\sigma$ of the Gaussian profile.
 \end{tablenotes}
\end{threeparttable}
\end{table*}

In Figure~\ref{fig04_opt}, we exhibit the distributions of $\sim$300 emission-line spectra color-coded by the BPT (panel a),  H$\alpha$ flux (panel b), H$\alpha$/H$\beta$ (panel c), and RS32-based gas-phase metallicity (12+log(O/H), panel d), respectively.  
Following the criteria of \cite{Kewley+01} and \cite{Kauffmann+03}, we employ the BPT diagnostic diagram to discriminate each spectrum in the central coverage by emission-line flux ratio, [NII]$\lambda6583$/H$\alpha$ and [OIII]$\lambda5007$/H$\beta$. In panel (a) of Figure~\ref{fig04_opt}, about 97\% of the central spectra are star-forming (in green), which is consistent with the results in \cite{Ge+21}. As expected, the northwest region with the WR features has the strongest H$\alpha$ flux. Compared to the outside of the WR region, the inside WR region has higher dust attenuation (i.e., higher H$\alpha$/H$\beta$ in panel c) and lower gas-phase metallicity (panel d). 

Adopting RS32-based gas-phase metallicity, \cite{Luo+21} defined the anomalously low-metallicity (hereafter ALM) region, where the oxygen abundance is significantly lower than the expected one. The expected $\rm (12+\log(O/H))_{exp}$ was derived by the empirical relation between 12+log(O/H) and local stellar surface mass density within a stellar mass bin (i.e., $\Sigma_*$-Z$_{\rm gas}$),  as shown by grey dots and curves in panel (a) of Figure~\ref{fig05_ALM}. $\rm \Delta \log(O/H)$ represented the gas-phase metallicity difference between the observed $\rm (12+\log(O/H))_{obs}$ and the expected $\rm (12+\log(O/H))_{exp}$. \cite{Luo+21} defined that for all star-forming spaxels in galaxies with stellar mass  $\rm M_* > 10^9 M_\odot$, spaxels outside 3$\times$ standard deviations ($\sigma_Z=0.037$) of the $\rm \Delta \log(O/H)$ Gaussian distribution (i.e., $\rm \Delta \log(O/H)<-0.111$) were regarded as the ALM region.
The fraction of ($M_*<10^9M_\odot$) dwarf galaxies in the SDSS-IV MaNGA survey (including $\sim$8000 unique galaxies) is $<$2\%, and those with $M_* < 10^{8.5} M_\odot$ account for less than 0.1\% (\citealt{Cano-Diaz+22}). 
Since the sample in \cite{Luo+21} was also from the SDSS-IV MaNGA survey, we can compare the $\rm \Delta \log(O/H)$ distribution of spaxels in galaxies with $M_*>10^9 M_\odot$ with that of spaxels in galaxies with $M_*<10^9 M_\odot$ if assuming that all dwarf galaxies behave like PGC 44685. Those 
 $\rm \Delta \log(O/H)$ values of spaxels in dwarf galaxies can be estimated by comparing the $\Sigma_*$-Z$_{\rm gas}$ relation within $8.5<\log M_*/M_\odot <9$ from \cite{Luo+21} since the ($<$0.1\%) fraction of galaxies with $M_* < 10^{8.5} M_\odot$ is too low to affect this relation. 
In panel (b) of Figure~\ref{fig05_ALM}, the solid grey curve shows the $\rm \Delta \log(O/H)$ distribution for dwarf galaxies. The dashed grey curve presents the $\rm \Delta \log(O/H)$ distribution in galaxies with $M_*>10^9 M_\odot$, derived from \cite{Luo+21}. The blue histogram is the total distribution of both. 
We fit only the positive side of the total metallicity deviation distribution to derive the symmetric negative side using a Gaussian profile, as done by \cite{Luo+21}. As expected, we get the same fitting result ($\sigma^{\prime}_Z=$0.037) as in \cite{Luo+21} where $\sigma_Z=$0.037, because dwarf galaxies do not contribute anything to the positive side of $\rm \Delta \log(O/H)$ distribution in this simple assumption. This suggests that the star-forming spaxels from dwarf galaxies could not significantly affect the ALM criterion ($\rm \Delta \log(O/H)<-3\sigma_Z=-0.111$) of \cite{Luo+21}. The orange vertical dashed line represents the location of -$3\sigma^\prime_Z$ (or -3$\sigma_Z$) below the center of Gaussian distribution.
Compared to this location,  we employ a stricter limit of -5$\sigma_Z$ (i.e., $\rm \Delta \log(O/H)< -0.185$, the red vertical dashed line) to classify the ALM region even for galaxies with $M_*<10^9M_\odot$ like PGC 44685, considering that the above assumption ignores the effects of the completeness and complexity of dwarf galaxies. 
We clarify that this applied stricter limit is obtained by using the fitting results (i.e., the Gaussian profile with $\sigma_Z=$0.037) from the sample of galaxies with $M_*>10^9M_\odot$.
The panel (a) of Figure~\ref{fig05_ALM} shows the $\Sigma_*$-Z$_{\rm gas}$ relation in these ALM spaxels by filled red circles. Open red and grey circles represent spaxles with $\rm \Delta \log(O/H)>-3\sigma_Z$ and $\rm -3\sigma_Z < \Delta \log(O/H) < -5\sigma_Z$, respectively. All these spaxels are corresponding to those in Figure~\ref{fig04_opt}. 
The ALM spaxels are $\lesssim$0.1 dex below the $\Sigma_*$-Z$_{\rm gas}$ relation in two low-mass bins of \cite{Luo+21}. The corresponding distribution of ALM spaxels in PGC 44685 is marked by white dots in panel (d) of Figure~\ref{fig04_opt}. Compared to the H$\alpha$ map, the ALM region reconciles with the region where H$\alpha$ is stronger.
In addition, we should note that these ALM spaxels may not actually be ``anomalous'' enough for low-mass galaxies, which requires further exploration in the future. The following analysis is based on the assumption that spaxels with $\rm \Delta \log(O/H)<-5\sigma_Z$  have anomalously low metallicity.

\begin{figure*}
\includegraphics[width=0.9\textwidth, angle=0 ]{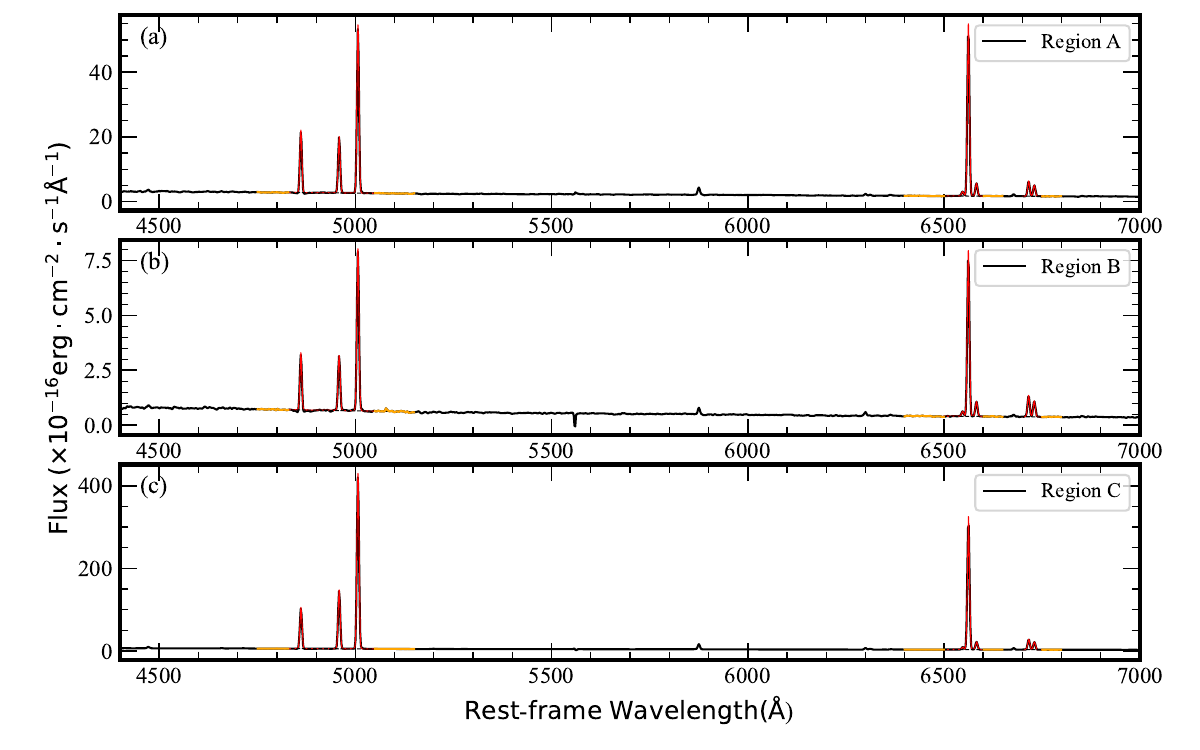}
\caption{Stacked anomalously low-metallicity (ALM) spectra in three star-forming regions. From the top to bottom panel, it shows the summed  ALM spectra in regions A, B, and C, respectively.  These star-forming regions are classified in section~\ref{sec3.3}.
The continuum and emission are fitted using the same method as in section~\ref{sec3.1}. } 
\label{fig06_3SF}
\end{figure*}

\begin{figure*}
\includegraphics[width=0.98\textwidth, angle=0 ]{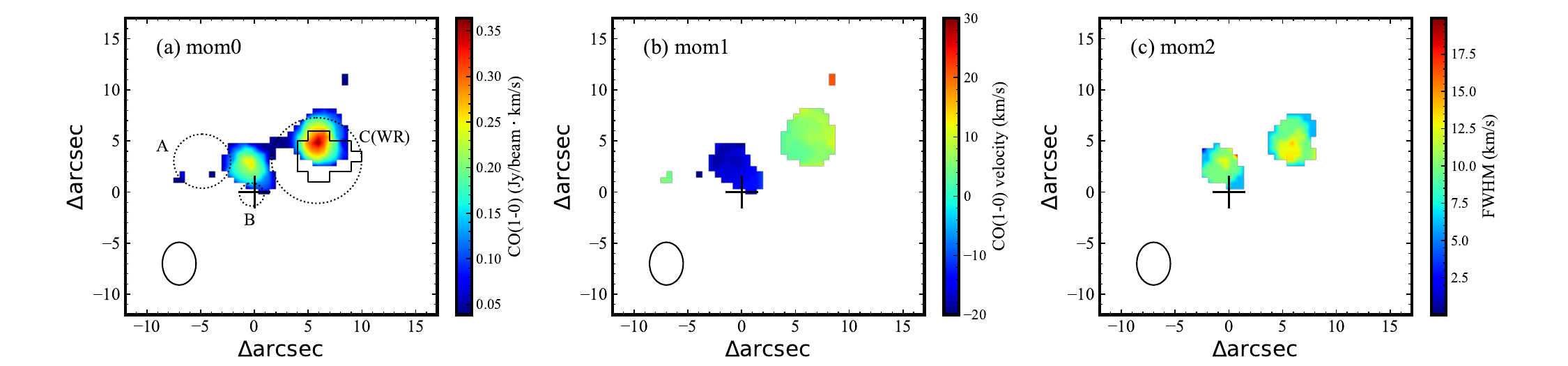}
\caption{Moment maps of the CO(1-0) emission with a threshold of $\rm >4\sigma_{rms}$.
The mom0 map in panel (a) is the integrated intensity, and mom1 (panel b) and mom2 (panel c) maps show the mean velocity and line width (FWHM) of CO(1-0), respectively. The symbols of the black plus and the bottom left ellipse indicate the center of the galaxy and the beam size (4.20\arcsec$\times$3.14\arcsec), respectively. The positions of triple star-forming (and WR) regions same as in Figure~\ref{fig04_opt} are marked in panel (a) by black dotted circles (solid polygon).  }
\label{fig07_mmt} 
\end{figure*}

\subsection{Triple star-forming regions} \label{sec3.3}

Since there are triple bright cores found in the star-forming region of PGC 44685, we can classify them into star-forming cores by using the python package {\it Astrodentro}\footnote{https://dendrograms.readthedocs.io/en/stable/} (\citealt{Goodman+09}) on the spatial resolved  H$\alpha$ distribution (see Figure~\ref{fig04_opt}). The hierarchical tree-diagram (i.e., `dendrogram') analysis in {\it Astrodentro} has already been used to identify star-forming cores by \cite{Li+20}. In {\it Astrodentro}, some parameters are needed to specify,  including {\it min\_value}, {\it min\_delta}, and {\it min\_npix}, for constraining the minimum value in regions, the minimum significance for structures and the minimum number of pixels in an independent structure, respectively. We adopted {\it min\_npix} =4, and manually increased 10\% flux around the galaxy center to better separate it from the northwest WR region.  Finally, we classify three star-forming cores/regions, 
marked by dotted circles and labelled with A, B, and C, respectively, in panels (b) and (d) of Figure~\ref{fig04_opt}.  

Figure~\ref{fig06_3SF} presents the stacked spectra that satisfy the ALM criteria in each star-forming region.
In region C, it covers a total of 55 spaxels, of which 54 belong to ALM spaxels. In region A, 20/22 spaxels are in the ALM region, but all (4/4) spaxels in central region B are ALM spaxels.  These spectra are dominated by the emission lines. Thus, we adopt the same approach as mentioned in section~\ref{sec3.1} to fit the continuum and emission lines. The best-fit Gaussians for  H$\alpha$, H$\beta$, [NII], and [SII] are displayed in red. The physical parameters in the three regions are listed in Table~\ref{tab3}. The ALM spaxels dominate in star-forming regions, so the physical parameters in ALM spaxels are slightly different from those considering all spaxels in the same star-forming region.

\begin{figure*}
\includegraphics[width=0.98\textwidth, angle=0 ]{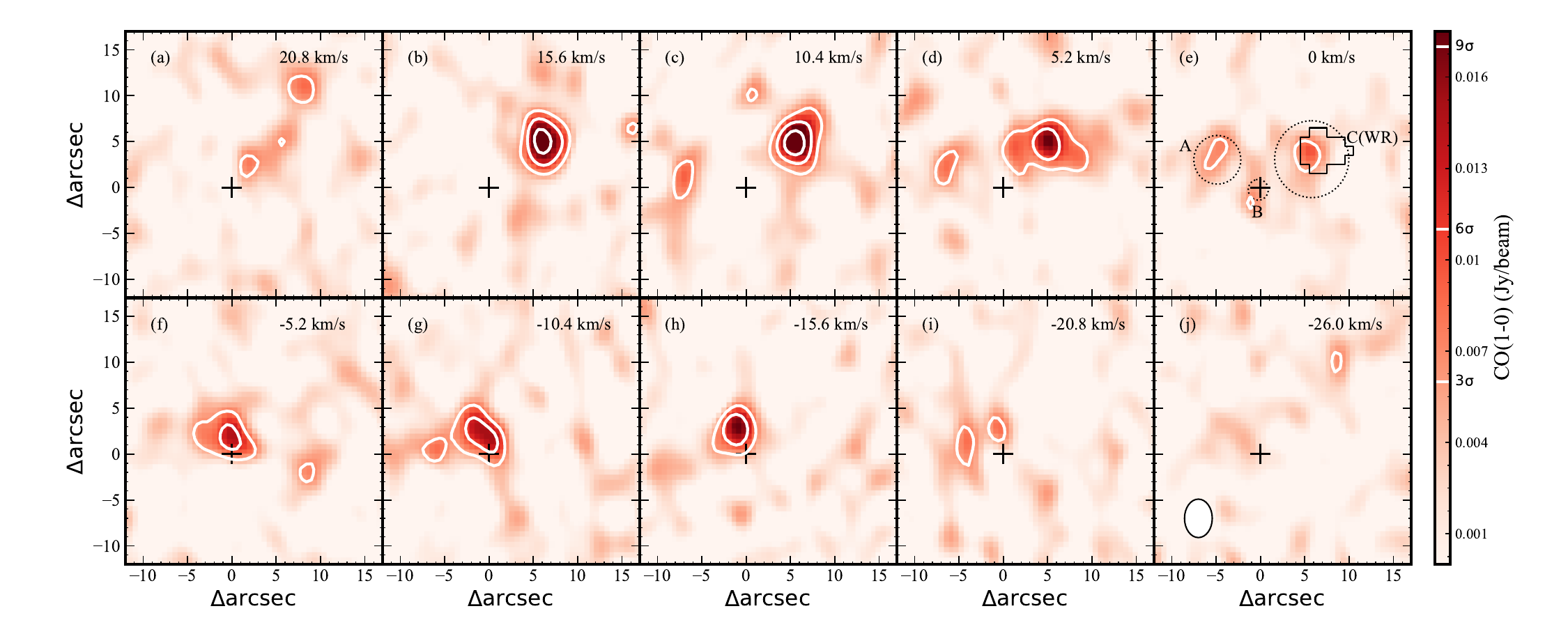}
\caption{Channel maps of the CO(1-0) emission with a velocity range of 20.8km/s $< v<$ -26.0km/s and a velocity interval $\rm \Delta$$v$ =5.2km/s. The corresponding value of the velocity is shown in the upper right corner. The galaxy center and beam size are marked with a black plus and a black ellipse, respectively.  The positions of triple star-forming (and WR) regions same as in Figure~\ref{fig04_opt} are marked in panel (e) by black dotted circles (solid polygon). The CO(1-0) intensity in white contours starts and increases at 3$\rm \sigma_{rms}$ ($\rm \sigma_{rms}=1.85$mJy/beam). The increments are marked in white on the right-side color bar. }
\label{fig08_channel}
\end{figure*}

\begin{figure}
\includegraphics[width=0.9\columnwidth]{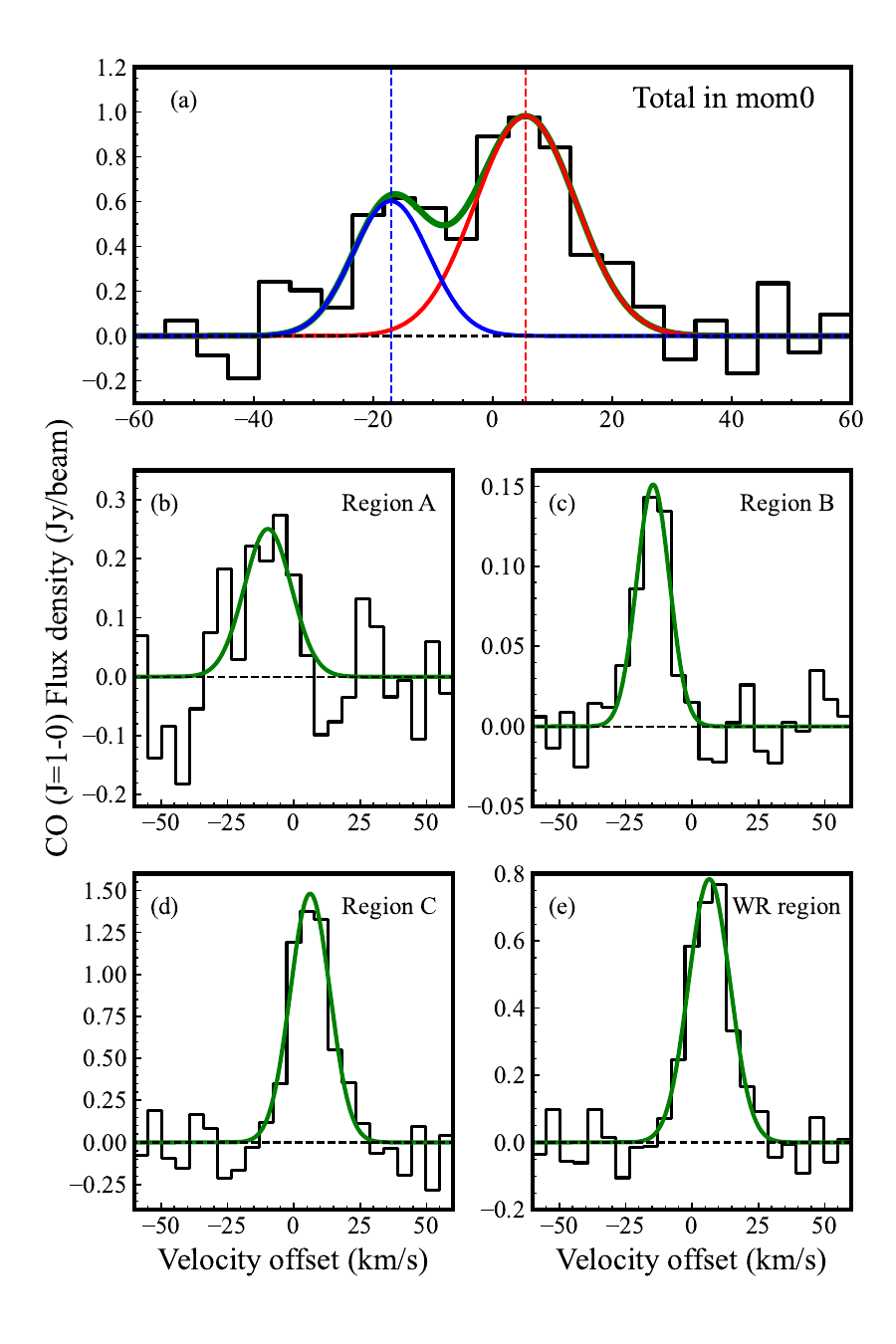}
\caption{The CO(1-0) line profile with a velocity interval of 5.2 km/s. In panel (a), the black histogram represents the original stacked spectrum within the mom0 map. We adopt Gaussian profiles to fit the redshift (in red) and the blueshift (in blue) parts of the line profile, in which corresponding centroids are shown in dashed lines. The green curve is the superposition of the two Gaussian profiles as the best-fit model. The spectra of pixels distributed in the ALM area are selected to be stacked in the same specific star-forming region. The corresponding stacked spectrum (black) and the best-fit single Gaussian profile (green) are shown in panels (b) to (e).  The region names are labelled at the top right of each panel. }
\label{fig09_profile}
\end{figure}

\subsection{NOEMA CO(1-0) distribution}\label{sec3.4}

Utilizing the MAPPING module in GILDAS, the CO(1-0) emission line and the underlying continuum could be separated. The moment maps of the CO(1-0) molecular line are present in Figure~\ref{fig07_mmt} including the flux intensity (panel a), the mean velocity (panel b), and the full width at half maximum (FWHM, panel c), respectively.  
The WR region contains the richest CO(1-0) molecular gas with the highest velocity dispersion as expected, while the other part of molecular gas distributes near the galaxy center (but mostly outside region B). 
The blueshift around the center means the molecular gas is moving toward the central region B. More snapshots of the velocity are shown in the channel maps of Figure~\ref{fig08_channel}. The intensity in each snapshot is highlighted by white contours in increments of 3$\rm \sigma_{rms}$, starting at 3$\rm \sigma_{rms}$. Panels (i) to (f) display the approaching side of molecular gas, while panels (d) to (a) show the receding side of molecular gas. In panel (e), the snapshot at zero velocity presents a small amount of CO(1-0) molecular gas in region A. These snapshots illustrate the kinematic connection among the triple bright regions in PGC 44685. Both the approaching and receding sides of the molecular gas cover a small range of velocity ($ | \Delta v |<30$ km/s), which is in the same magnitude as the molecular gas velocity surrounding  the WR stars in the inner Galactic plane (\citealt{Baug+19})

\begin{figure*}
\includegraphics[width=0.9\textwidth, angle=0 ]{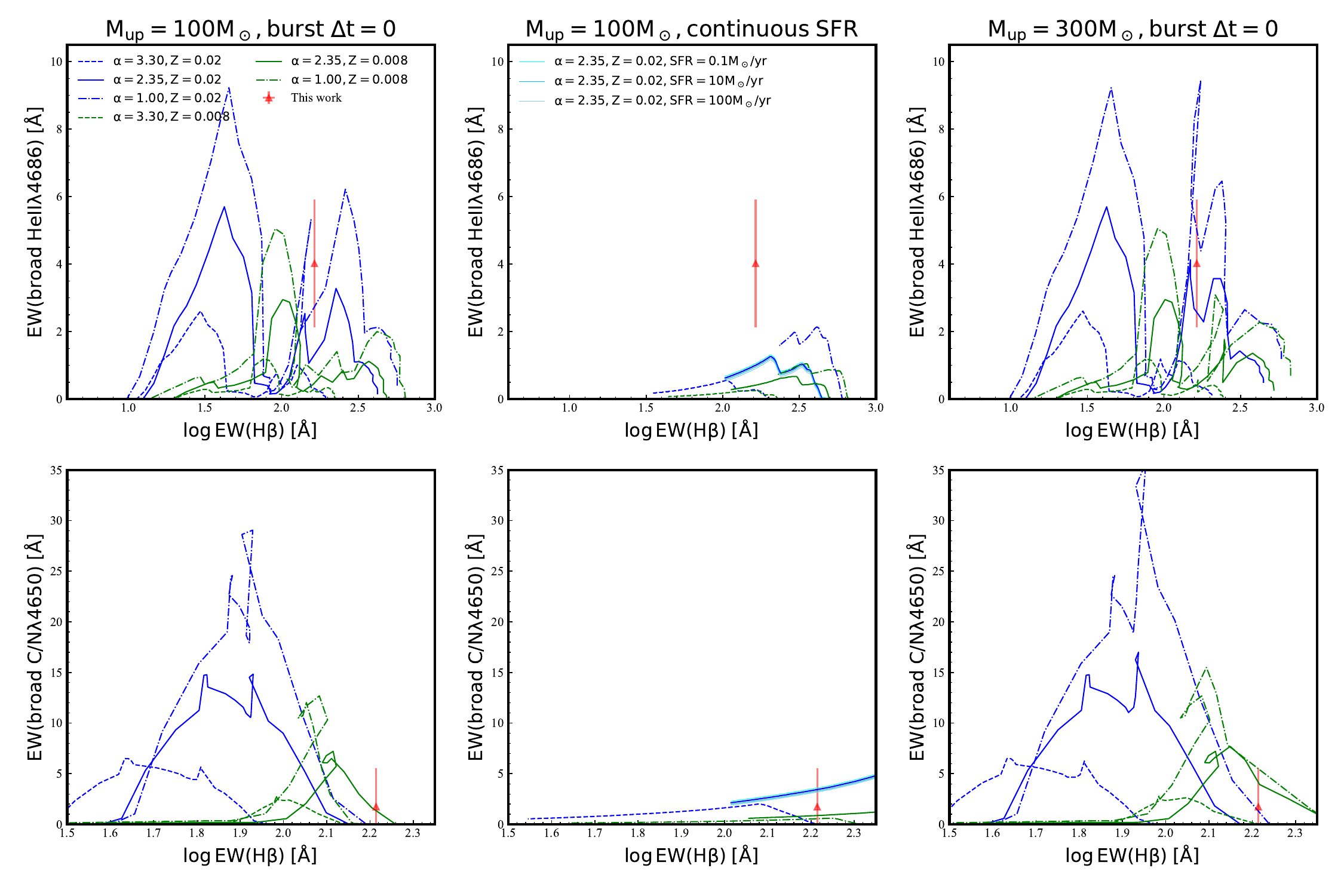}
\caption{ Distributions of the equivalent widths of the broad HeII$\lambda$4686 (top panels) and C/N$\lambda4650$ (bottom panels) bumps vs. that of H$\beta$. The simulated evolutionary models from \textit{Starburst99} are divided into two metallicities, i.e., $Z$=0.02 (solar, in blue) and 0.008 (sub-solar, in green), and three IMF slopes, $\rm \alpha=$3.30 (steeper slope,  in dashed line), 2.35 (Salpeter slope, in solid line), and 1.00 (flatter slope, in dash-dot line). The red triangle marks the result of this work, and the corresponding errors are estimated by the method of \cite{Cayrel+1988}. 
The title of each top panel expresses the adopted burst duration ($\Delta t$) and the upper cutoff mass ($M_{\rm up}$) in each case. In the middle panels, we additionally consider three extended burst models with durations $\Delta t$= 0.2 (SFR=100 $M_\odot$/yr), 2 (10$M_\odot$/yr), and 200 Myr (0.1$M_\odot$/yr) in the case of Salpeter IMF ($\alpha$=2.35), shown in different blue-like colors (moved slightly up/down for clarity). The top panels (bottom panels) share the same range of axes for visual comparison.}
\label{fig10_SB99}
\end{figure*}

The decomposition of the CO(1-0) line profile within the mom0 map is displayed in panel (a) of Figure~\ref{fig09_profile}. The red and blue Gaussian components correspond to the receding and approaching sides of CO(1-0) molecular gas, respectively. The best-fit CO(1-0) line profile, therefore, exhibits a bimodal shape. Based on this line profile, we can estimate the CO line luminosity by adopting the relation in \cite{Solomon+05}, 
\begin{equation}
\label{eq8}
   L'_{\rm CO} [{\rm K\; km\; s^{-1}}]= 3.25\times 10^{7} S_{\rm CO}\Delta v \nu_{\rm obs}^{-2} D_{L}^{2} (1+z)^{-3},
\end{equation}
where $S_{\rm CO}\Delta v$, $\nu_{\rm obs}$, and $D_{\rm L}$ are the CO integrated flux density in the unit of $\rm Jy \, km \, s^{-1}$, the observed frequency in GHz, and the luminosity distance in Mpc at a given redshift, respectively. Assuming the conversion factor ($\rm \alpha_{CO}$) between CO(1-0) and $\rm H_2$ of 4.3 $\rm (M_\odot km\; s^{-1} \;pc^2)^{-1}$ in \cite{Bolatto+13}, this corresponding mass of the molecular hydrogen ($\rm M_{ H_2}$) can be assessed by :
\begin{equation}
\label{eq9}
    {\rm M_{H_2}} = \alpha_{\rm CO} \times L'_{\rm CO}.
\end{equation}
Both total $L^{'}_{\rm CO}$ and $M_{\rm H_2}$  in the galaxy are listed in Table~\ref{tab1}. 
Panels (b) to (e) of Figure~\ref{fig09_profile} show the stacked spectra in the ALM distributions of regions A, B, C, and WR, respectively. 
We adopt a single Gaussian profile to fit each of them, and the best fits are shown in green. The estimated $L^{'}_{\rm CO}$ and $M_{\rm H_2}$  in a specific region are listed in Tables~\ref{tab2} and \ref{tab3}. 
About 90\% (50\%) of the total $M_{\rm H_2}$ is located within region C (WR region). The WR region has the highest molecular gas density with respect to other regions or the entire galaxy (see Tables~\ref{tab1}, \ref{tab2} and \ref{tab3}). The total $M_{\rm H_2}$ estimation in moment maps is smaller than that in \cite{Ge+21}, because they estimated it within a larger coverage of the IRAM 30-m observation.
Here, the $M_{\rm H_2}$ could be considered a lower limit due to the presence of abundant atomic gas ($\rm M_{HI} \sim 10^{8.27} M_\odot$, \citealt{Haynes+18}).

\section{Discussions} \label{sec4-discussions}

\subsection{WR spectral constraints on the IMF slope and metallicity} \label{sec4.1}

The WR galaxies are a rare population due to the small number and short lifetime of WR stars. 
PGC 44685 is a complexity of multiple types of galaxy classification and hosts WR stars in its most active star-forming region. 
In this study, we classified PGC 44685 as a dSFS0, which allows us to explore the constraints of WR features on the IMF slope and the metallicity roughly. Hence, we adopt the published online stellar evolutionary model, \textit{Starburst99} (hereafter, SB99, \citealt{Leitherer+Heckman+1995, Leitherer+1999}), to simulate the evolution of the WR population. Although it's designed for the starbursts or a young stellar population, it also incorporates the WR population in some models, detailed WR features described in \cite{Schaerer+Vacca+1998} and \cite{Garcia-Benito+10}. Each model mainly includes four input parameters: the burst duration ($\Delta t$), the total mass of stars formed in the burst ($\rm M_{total}^{SB99}$), the IMF (i.e., a power low $\rm dN/dM \propto M^{-\alpha}$ with slope $\alpha$), and the metallicity ($Z$). Since the WR population is very young and short-lived, we mainly employ the instantaneous star formation as commonly studied in the literature (\citealt{Zhang+07, Liang+21}), i.e., $\Delta t=0$, and set $\rm M_{total}^{SB99}=1.0\times 10^6 M_\odot$, which is only used as a normalization constant (\citealt{Schaerer+Vacca+1998}). Considering that the WR phase is accompanied by strong stellar winds, we select the Geneva tracks with ``high" mass-loss rates.  For the IMF, we adopt the low and up boundaries of $M_{\rm low}=0.1 M_\odot$ and $M_{\rm up}=100 M_\odot$, with three different slopes (a flatter slope $\alpha=1.0$, a Salpeter slope $\alpha=2.35$, and a steeper slope $\alpha=3.3$). Due to the dependence of IMF on the metallicity (e.g., \citealt{Liang+21}), we try two metallicities, i.e., one is the solar metallicity $Z$=0.02, and the other is the sub-solar metallicity $Z$=0.008, considering that the Geneva tracks have not provided finer grids of metallicity. The time grid is from 0.1 Myr to 20 Myr with an increment of 0.1 Myr. Based on the above assumptions, the SB99 predictions are shown in the left panels of Figure~\ref{fig10_SB99}. 

In addition, the possible dependences on the burst duration ($\Delta t$) and the upper cutoff mass ($M_{\rm up}$) are tested as well. For the burst duration ($\Delta t$), we adopt the continuous star formation (i.e., SFR=1$M_\odot$/yr and $\Delta t=20$Myr) with $M_{\rm up}=100M_\odot$ in each metallicity and IMF slope. Following  \cite{Zhang+07} and \cite{Liang+21}, we also consider in the case of Salpeter IMF ($\alpha=2.35$) three additional extended burst models with durations $\Delta t=$ 0.2, 2, and 200 Myr. These SB99 predictions assuming the continuous star formations are shown in the middle panels of Figure~\ref{fig10_SB99}.  At the fixed Salpeter IMF with solar metallicity, adopting different SFRs, shown as a set of light blue, can not affect the distribution in the spaces of EW(HeII$\lambda4686$) and EW(C/N$\lambda4650$) vs. EW(H$\beta$).  All of the continuous star formations with different metallicity and IMF slopes (in blue and green) only make a few contributions to the EW(HeII$\lambda4686$) and EW(C/N$\lambda4650$). 
For the $M_{\rm up}$, we replace it with a large value of 300 $M_\odot$, also adopted by \cite{Liang+21}, but still use an instantaneous burst model with the same other parameters setting,  corresponding results shown in the right panels of Figure~\ref{fig10_SB99}.

Figure~\ref{fig10_SB99} exhibits these predictions of the WR population in the space of the equivalent widths (EW) of broad bumps (HeII$\lambda4686$ in the top panels and C/N$\lambda4650$ in the bottom panels) and H$\beta$. In an instantaneous burst, the EW(H$\beta$) can be served as an age indicator of the population (\citealt{Copetti+1986}). The solar and sub-solar metallicities are present in blue and green lines, respectively. The steeper, Salpeter, and flatter slopes are in the dash, solid, and dash-dotted lines separately. 
In the spaces of EW(HeII$\lambda4686$) vs. EW(H$\beta$) at top panels,  the properties of WR features in PGC 44685 (in a red triangle) seemingly reconcile to the prediction with IMF $\alpha$=1.0 / 2.35 and $Z$=0.02, while in the EW(C/N$\lambda4650$) vs. EW(H$\beta$) distributions at the bottom panels, our results trend to agree with the prediction with IMF $\alpha $= 1.0 / 2.35 and $Z$=0.008 / 0.02. There is no denying that our result gives a large uncertainty due to the only one target with large error bars. We still could contribute to the degeneracy between IMF and metallicity. Our results could roughly hint that the WR properties are possibly preferred to reconcile with the IMF slope $\alpha \le $2.35 and the metallicity $Z \le 0.02$, which is consistent with the previous studies (\citealt{Zhang+07, Liang+21}). But this result still needs a large sample and simulations with a finer grid to verify the robustness in the future.

\begin{figure*}
\includegraphics[width=0.98\textwidth, angle=0 ]{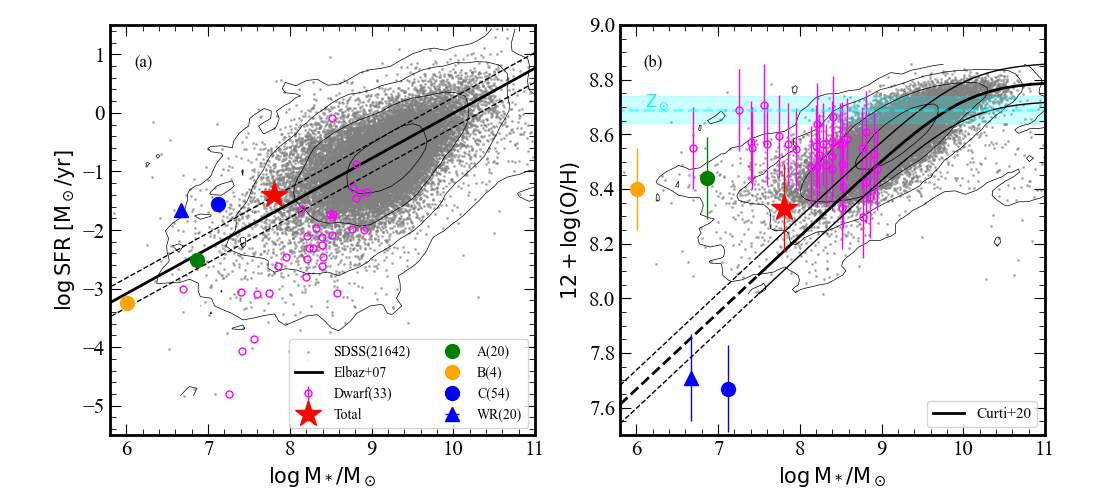}
\caption{Star-forming main sequence (SFMS, panel a) and mass-metallicity relation (MZR, panel b).  The red star represents the total galaxy, while other specific regions with ALM-spaxel numbers are plotted in different shapes and colors. The grey contours and dots present the sample of 21642 galaxies from the NASA-Sloan Atlas catalog (\citealt{Blanton+11}). The magenta circles are 33 local dwarf galaxies from the Spitzer Local Volume Legacy (LVL, \citealt{Dale+09}) survey, which were selected by Li et al. (in preparation) with spatially resolved observations. In panel (a), the SFMS of local star-forming galaxies is derived by \cite{Elbaz+07}, shown in the black solid and dashed lines. In panel (b), the MZR in the local universe is from \cite{Curti+20} by adopting different metallicity calibrations, shown in solid lines, while the MZR with stellar mass $\rm M_* < 10^{7.5}M_\odot$ is extrapolated, depicted by black dashed lines.  The dashed cyan line with the shaded region marks the value of solar abundance (i.e., $\rm Z_\odot=8.69\pm0.05$, \citealt{Allende+Prieto+01}).}  
\label{fig11_MS_MZR}
\end{figure*}

\subsection{SFMS and MZR in different regions} \label{sec4.2}
We further explore the star-forming main sequence, i.e., SFMS, and the stellar mass-metallicity relation, i.e., MZR, for galaxy PGC 44685 in different specific regions, as shown in Figure~\ref{fig11_MS_MZR}. Only those spaxels that satisfy the ALM criteria are considered in a specific region.
The sample of 21642 galaxies in the local universe is selected from the NASA-Sloan Atlas Catalog\footnote{http://nsatlas.org/data} (\citealt{Blanton+11}) by limiting the emission line with SNR $>10$ and removing some unreasonable results, which are shown in grey dots and contours. For the local dwarf sample,  Li et al. (in preparation) selected a total of 65 galaxies with $10^6 M_\odot <M_*<10^9 M_\odot$ from the Spitzer Local Volume Legacy Survey (LVL, \citealt{Dale+09}). Subsequently, they are conducting the Dwarf Galaxies Integral Survey (DGIS) to acquire the corresponding spatially resolved optical data (with the Multi Unit Spectroscopic Explorer of the Very Large Telescope, MUSE/VLT,  and the Wide Field Spectrograph of the Australian 2.3-m telescope, WiFeS/ANU). Based on this dwarf sample, we select 33 dwarf galaxies with the SNR of emission lines larger than seven, shown in magenta circles. The black lines in panel (a) present the SFMS of local star-forming galaxies from \cite{Elbaz+07}, while the solid lines in panel (b) are the local MZR in \cite{Curti+20}, which are extrapolated to lower stellar mass ($<10^{7.5} M_\odot$), shown in dashed lines. 
The MZR in \cite{Curti+20} was derived by adopting a combination of metallicity diagnostics, including R3 (=[OIII]$\lambda$5007/H$\beta$), R2 (=[OII]$\lambda\lambda3727,3729$/H$\beta$), N2 (=[NII]$\lambda$6584/H$\alpha$), S2 (=[SII]$\lambda\lambda$6717,6731/H$\alpha$), O3O2 (=[OIII]$\lambda$5007/[OII]$\lambda\lambda$3727,3729) and O3N2 methods. 
The above parameters in Figure~\ref{fig11_MS_MZR} are compiled by the same IMF and RS32-based method as in this work, except for the stellar mass of dwarf galaxies, which was derived from Spitzer 3.6$\mu$m with a mass-to-light ratio by Li et al. (in preparation), and the metallicity of the local MZR in \cite{Curti+20}.

In Figure~\ref{fig11_MS_MZR}, we find that the total galaxy PGC 44685 follows the local SFMS and MZR within the error, and places within the scatter of dwarf galaxies.  The star formation in region C (WR) is 0.2 (0.3) dex higher than the main sequence, but the gas-phase metallicity is lower than the extrapolation of local MZR in \cite{Curti+20}. The star formation activities in regions A and B still follow the SFMS, and their metallicity is somewhat comparable to that of some local dwarf galaxies, but $\sim$0.6 dex higher than the extrapolation of the local MZR in \cite{Curti+20}. The difference between regions A/B and C/WR could probably be relative to the strong star formation activity indicated by the presence of the WR population. 
Because during the short lifetime ($\sim$ 2-5 Myr) of WR stars, the most massive O stars have evolved, and WR stars may release more nitrogen into the local interstellar medium (ISM, \citealt{Berg+11, Perez-Montero+11, Monreal-Ibero+12, Maiolino+Mannucci+19}), and oxygen could be blown out more preferentially by (possibly) WR-driven stellar winds ($\sim$200 km/s) than nitrogen (\citealt{Maiolino+Mannucci+19, Roy+21}). 
This is consistent with what we found in section~\ref{sec3.1}, where the possible stellar winds driven by WR stars are $<$500 km/s, and WNL stars are dominated during the WR phase. More WNL stars suggest that more nitrogen-rich WR stars exist. Since the CAHA optical data does not cover [OII]$\lambda \lambda 3727,3729$, we can not directly confirm the nitrogen enhancement by N$^+$/O$^+$ ratio.
Additionally, the different metallicity between different regions can reflect the fact that the star formation and evolution of galaxy PGC 44685 is relatively complex because both the level of star formation and gas flows (accretion or outflow) can affect and regulate the metal content (\citealt{Maiolino+Mannucci+19}). We should also note that the MZR at the low-mass side (especially for $< 10^{7.5} M_\odot$) is still little known (\citealt{Maiolino+Mannucci+19, Curti+20}), so one thing can be confirmed that the gas-phase metallicity in regions A, B, and C(WR) or in the total galaxy is smaller than the solar abundance ($\rm Z_\odot=8.69\pm0.05$, \citealt{Allende+Prieto+01}), as shown in cyan. We will discuss the possible scenario of the star formation in different regions in section~\ref{sec4.4}. 

\begin{figure*}
\includegraphics[width=0.98\textwidth, angle=0 ]{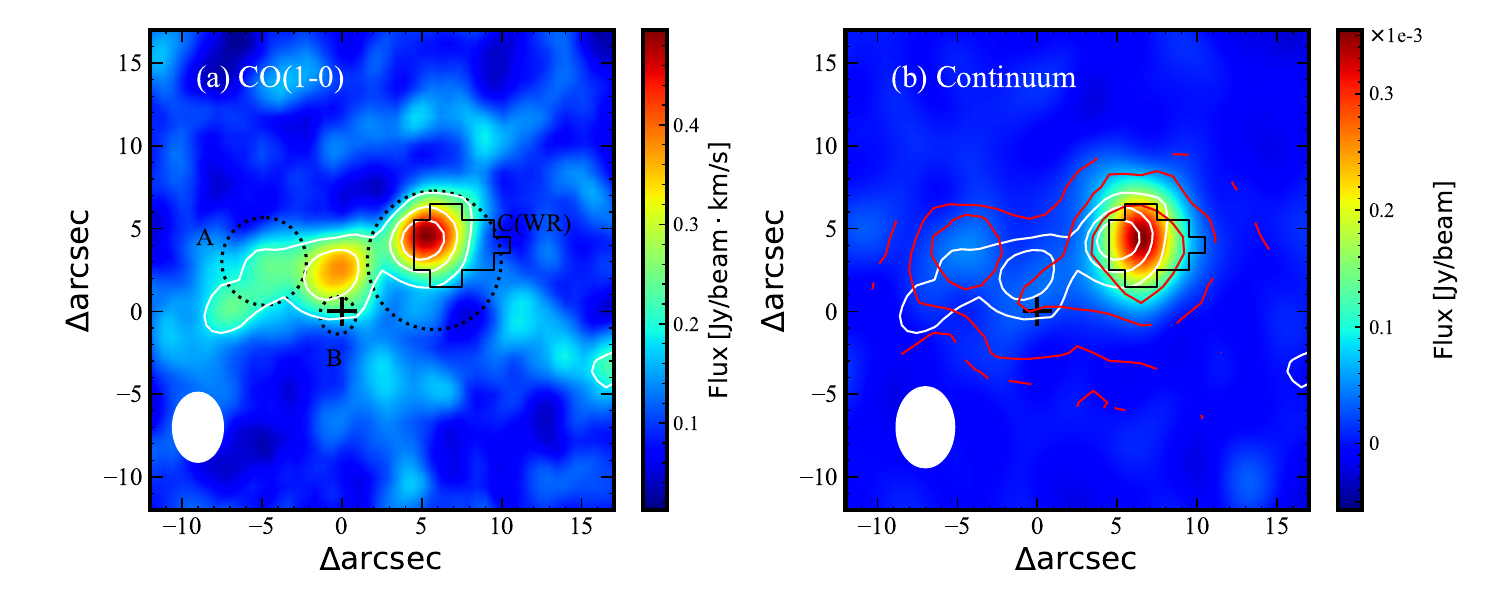}
\caption{ Integrated intensity maps of CO(1-0) emission (panel a) and 3-mm continuum (panel b). The white contours represent the CO(1-0) intensity at levels of 0.2, 0.3, and 0.4 Jy/beam$\cdot$km/s, respectively. The black polygons and circles show the WR region and three star-forming regions, respectively. The red contours in panel (b) correspond to the distribution of H$\alpha$ flux in Figure~\ref{fig04_opt} at the 5\%, 10\%, and 30\% of the peak flux. Black plus marks the galaxy center. The beam sizes of CO(1-0) emission ($\rm 4.2\arcsec \times 3.14\arcsec$) and continuum ($\rm 4.84\arcsec\times 3.61\arcsec$) are plotted in white ellipses in the left-bottom of each panel.  }
\label{fig12_gas}
\end{figure*}

\subsection{H$\alpha$, CO(1-0) and 3mm continuum distributions} \label{sec4.3}
A large fraction of stellar radiation from ultraviolet in galaxies is absorbed by dust, which will be re-radiated at the infrared wavelengths. Therefore, the millimeter distribution can trace the obscured region with active star formation. Thanks to the deep NOEMA observation ($\sim$9.3 hrs) of the 3mm band on source PGC 44685, we could make it possible to explore the distributions of both CO(1-0) molecular gas and the corresponding 3mm dust continuum, simultaneously. Since the 3mm flux intensity comes from the faint tail of the Rayleigh-Jeans regime of the dust thermal emission, its intensity is usually several times fainter than that at shorter wavelengths, such as the sub-mm band $\lambda \sim 850$-1100$\mu$m. Furthermore, there is a possibility that part of the 3mm dust continuum could have a non-thermal origin associated with the WR winds.

Figure~\ref{fig12_gas} plots the integrated intensity distributions of CO(1-0) emission (panel a) and 3mm dust continuum (panel b). Triple star-forming regions and WR region are overlaid on panel (a), and the H$\alpha$ flux distribution is displayed in red contours on panel (b). Although the continuum intensity is much weaker than that of molecular gas, the detection of dust continuum hints at the high star formation activity with high dust obscuration in the WR region. As expected, the region C(WR) has the highest molecular gas reservoir, the strongest dust continuum, and the highest SFR (see Tables~\ref{tab2}, \ref{tab3}, and Figures~\ref{fig04_opt}, \ref{fig11_MS_MZR}). In contrast, the CO gas fraction and H$\alpha$-based SFR within region A are lower than region C(WR), but slightly larger than in region B within the error. If we assume a constant star formation rate and do not consider the supply of atomic gas, the host galaxy (and region C or WR) could quickly consume the remaining molecular gas in $<$ 30 Myr (36 or 25 Myr) and become quiescent. This could be a lower limit on gas consumption when considering conversions between atomic and molecular gases.
Since the equivalent width of H$\beta$, i.e., EW(H$\beta$), decreases with time within a given star-forming region, it has been considered as a gross tracer of the age, decreasing as the HII nebular gets older (\citealt{Dottori+1981, Terlevich+04, Cuisinier+06}).  Tables~\ref{tab2} and \ref{tab3} list the EW(H$\beta$) in the WR region and star-forming regions A, B, and C, respectively. If we don't consider the scatter induced by the accumulation of old populations from previous star formation activities,  region C (or WR) with the higher EW(H$\beta$) has a younger age, then followed by region A. The age in region B is the oldest due to the lowest EW(H$\beta$). If combining this informative EW(H$\beta$) with CO gas and SFR in different regions, it seems to suggest that the star formation triggered by the gas reservoir in regions A/B is before that in region C(WR), which is the reason for few gases left in regions A/B.  

Additionally, we also notice that in region C (see panel b of Figure~\ref{fig12_gas}),  the 3mm dust continuum plateau reconciles with the H$\alpha$ flux peak (red contours), but it is slightly shifted westward by about 2\arcsec against the local molecular gas (white contours). We deduce that this slight eastward shift of molecular gas may be caused by gas accretions or the relatively weaker stellar winds (i.e., $<$500 km/s) accompanied by the WR phase. \cite{Ge+21} proposed that the significant rotation was observed near the WR region of PGC 44685, where there is low-velocity dispersion of stars and ionized gas alongside high star formation.
\cite{Liang+21} pointed out that more WR stars (i.e., a top-heavy IMF) at low metallicity can compensate for their weaker stellar winds. 
Based on our current optical spectra (spectral resolution of $\sim$150 km/s), we can not clearly resolve the tentative outflow component from the optical ionized emission lines (such as [OIII]$\lambda \lambda4959,5007$)  surrounding the WR region. Thus, we have proposed a GTC/MEGARA observation with a higher spectral resolution ($\sim$ 25 km/s) to attempt to detect the outflows driven by the theoretical stellar winds. A full physical analysis of the gas will also be available in the next paper.

\subsection{ Gas accretion inducing star formation? }\label{sec4.4}
The center of dSFS0 PGC 44685 contains the triple bright core, as shown in Figure~\ref{fig02_image}. Region C with the youngest age has the highest SFR and molecular gas, dominated by the WR population, while region B with the oldest age has the lowest SFR and gas. Region A is placed between these two (see Tables~\ref{tab2} and \ref{tab3}). Why does it have such distributions of different materials, and how does it form and evolve? We do not find any major merger remnants left around this galaxy. \cite{Xiao+16} have shown that this galaxy is in a sparse environment, meaning that the probability of a major merger disrupting the central part of the galaxy and triggering star formation is very low. \cite{Ge+21} suggested that the star formation of dSFS0s may originate from a long dynamic time-scale evolution in a low-density environment. Therefore, we speculate that the segmented distributions of molecular gas might be due to the gas-rich accretion from the ISM or circumgalactic medium (CGM), leading to the complexity of star formation subsequently. The gas supplement by accretions provides the raw material for star formation, then sustains it to follow the SFMS and MZR, as shown in Figure~\ref{fig11_MS_MZR}. If considering the contribution of atomic gas, \cite{Ge+21} has proved that the total galaxy PGC 44685 also followed the Kennicutt–Schmidt law as normal star-forming galaxies in the local universe.

As mentioned in section~\ref{sec3.2}, we adopt a stricter criterion to select spaxels belonging to the ALM region (see Figures~\ref{fig04_opt} and \ref{fig05_ALM}), where spaxels are assumed to have anomalously lower metallicity in the low-mass galaxies. \cite{Luo+21} speculated that ALM regions are the site of recent accretion of low-metallicity gas, which mixed with pre-existing metal-rich gas and triggered the local star formation.  In panels (b) and  (d) of Figure~\ref{fig04_opt}, we have found that the region with high H$\alpha$ intensity has lower gas-phase metallicity, which is consistent with the speculation of \cite{Luo+21}. Therefore, we speculate a possible scenario that the gas in PGC 44685 might be from the (minor) metal-poor gas accretion, which supports the raw material for star formation. The star formation in regions A/B could occur before that of region C(WR) because regions A/B have older age and less gas. The relatively longer past star formation could produce more heavy elements, leading to the larger gas-phase metallicity compared to the WR region. In region C(WR), the existence of the WR population would be the other reason for the lowest metallicity, because WR stars could dilute the oxygen abundance by possible stellar winds and produce more nitrogen (\citealt{Maiolino+Mannucci+19, Roy+21}).

Different from ETGs or normal SFS0s with $M_*>10^9 M_\odot $, molecular gas fragments and low stellar mass of PGC 44685 suggest that dSFS0s may undergo complex star formation activities, and the star formation might depend on the stellar mass. As reported cases in \cite{Ge+21}, when we compare images of dSFS0s, including PGC 44685, to those of normal SFS0 with stellar mass $M_*>10^9 M_\odot $ in the literature (\citealt{Xiao+16, Ge+20, Chen+21, Lu+22}), it seems that these dwarf ($M_*<10^9 M_\odot $) SFS0s show more clumpy star-forming cores or knots. \cite{Alatalo+13} reported the Combined Array for Research in Millimeter Astronomy (CARMA) ATLAS$\rm ^{3D}$ molecular gas imaging survey in CO-rich ETGs with stellar mass $M_*>6\times10^9 M_\odot $, and found that most massive galaxies have a disc molecular gas.  \cite{Fraser-McKelvie+18b} pointed out a mass-dependence scenario that massive ($M_*>10^{10} M_\odot $) S0s with older bulges than discs may go through the morphological or inside-out quenching while less massive  ($M_*<10^{10} M_\odot $) ones undergo the disc fading. 
It is not yet clear whether dSFS0s follow a similar quenching process or host a rotating stellar disk. Therefore, the question of their nature as true dwarf S0s and their connection to more massive classical S0s is far from being resolved. More statistical analysis is needed to confirm.
But it would be difficult and expensive because the long-exposure observations are necessary to detect the fragmented molecular gas in dwarf galaxies. In the WR galaxy catalog from MaNGA in \cite{Liang+20}, only $<$2\% of WR galaxies have the early-type morphologies and less massive stellar mass ($M_*<10^9 M_\odot $), simultaneously. \cite{Paudel+23} presented an extensive catalog of 5405 early-type dwarf (dE) galaxies in the various environments (a total sky area of 7643 deg$^2$) of the local universe (z$<$0.01) and also pointed out the rarity of dEs in a field, about 1.47 dEs per field. The rare case in this work thus could provide a chance to glance at the formation and evolution of dwarf lenticular galaxies (or analogues) with star formations and make attempts to accumulate more cases for future work.

\section{Summary} \label{sec5-summary}
Combining the 2D optical spectroscopy with a new millimeter observation of NOEMA, we study a dwarf lenticular galaxy PGC 44685 with triple star-forming regions (i.e., regions A, B, and C). In region C, we clearly detect the spectral characteristics of the WR population, implying an ongoing star formation activity. A dwarf galaxy with WR features is relatively rare due to the paucity and short lifetime of WR stars. Given spatially resolved observations, we attempt to figure out the trigger and evolution of star formation in the galaxy with a complex structure. Our main results are as follows:
\begin{itemize}
    \item Based on the spectral characteristics in each spaxel, we identify 20 spectra with WR features in region C. By fitting the stacked spectrum within this WR region, we can estimate that there are tens of ($<$30) WNL, few WCE ($<$3) stars and hundreds of ($<$300) O stars in the WR population. The stage of the violent star-forming region with higher dust may be in the nitrogen-rich WR phase, reconciling to the local low gas-phase metallicity. 
   \item The detection of CO(1-0) molecular gas with NOEMA reveals that most gas reservoir is mainly distributed in the C(WR) region, and less gas in regions A/B. Except for the CO(1-0) emission line, we also detect that the 3mm dust continuum distribution concentrates on the WR region.  Compared to the flux plateaus of H$\alpha$ and 3mm continuum in the WR region, the flux peak of CO(1-0) molecular gas is slightly shifted eastward by $\sim$2\arcsec. This might be affected by the gas accretion or the stellar winds with a velocity of hundreds of kilometers per second (i.e., $<$500 km/s), accompanied by the WR phase.
   \item  Considering the physical properties in regions A, B, and C(WR), we speculate that the star formations in region A/B are before that in region C(WR), because less CO gas, lower SFR and older age are found in regions A/B. 
Assuming that the stricter criterion of the anomalously low-metallicity (ALM) region could be applied to the low-mass galaxies, we found that most ($\sim$96\%) spaxels in regions A, B and C are in ALM region, suggesting the possible (minor) accretions of metal-poor gas providing the raw material for star formation. For the total galaxy, it still follows the SFMS and MZR relations. The galaxy would become quiescent by depleting the remaining gas after tens ($<$30) of Myr if there is no new supply of raw material for star formation. The complexity of star formation and clumpy molecular gas distributions in this dSFS0 case may suggest the complex star formation for SFS0s in the field, which needs further exploration. 
\end{itemize}

\acknowledgments
The authors are very grateful to the anonymous referee for critical comments and instructive suggestions, which significantly strengthened the analyses in this work.
This work is supported by the National Natural Science Foundation of China (No. 12121003, 12192220, and 12192222). We also acknowledge the science research grants from the China Manned Space Project with NO. CMS-CSST-2021-A05. Y.L.G acknowledges the grant from the National Natural Science Foundation of China (No. 12103023). RGB acknowledges financial support from grants CEX2021-001131-S funded by MCIN/AEI/10.13039/ 501100011033 and PID2022-141755NB-I00. This work is based on observations carried out with the IRAM Northern Extended Millimeter Array. IRAM is supported by INSU/CNRS (France), MPG (Germany), and IGN (Spain). In addition, we acknowledge the support of the staff from CAHA and NOEMA, especially Edwige CHAPILLON. 

\software {GILDAS (\citealt{Pety+05,Gildas+Team+13}), pPXF (\citealt{Cappellari+Emsellem+04, Cappellari+17}), MPFIT (\citealt{Markwardt+09}), Astrodendro (\citealt{Goodman+09}), $Starburst99$ (\citealt{Leitherer+Heckman+1995, Leitherer+1999}), GALFIT (\citealt{Peng+02}) }

\bibliography{ms}{}

\begin{thebibliography}{}
\expandafter\ifx\csname natexlab\endcsname\relax\def\natexlab#1{#1}\fi
\providecommand{\url}[1]{\href{#1}{#1}}
\providecommand{\dodoi}[1]{doi:~\href{http://doi.org/#1}{\nolinkurl{#1}}}
\providecommand{\doeprint}[1]{\href{http://ascl.net/#1}{\nolinkurl{http://ascl.net/#1}}}
\providecommand{\doarXiv}[1]{\href{https://arxiv.org/abs/#1}{\nolinkurl{https://arxiv.org/abs/#1}}}

\bibitem[{{Alatalo} {et~al.}(2013){Alatalo}, {Davis}, {Bureau}, {Young},
  {Blitz}, {Crocker}, {Bayet}, {Bois}, {Bournaud}, {Cappellari}, {Davies}, {de
  Zeeuw}, {Duc}, {Emsellem}, {Khochfar}, {Krajnovi{\'c}}, {Kuntschner},
  {Lablanche}, {Morganti}, {McDermid}, {Naab}, {Oosterloo}, {Sarzi}, {Scott},
  {Serra}, \& {Weijmans}}]{Alatalo+13}
{Alatalo}, K., {Davis}, T.~A., {Bureau}, M., {et~al.} 2013, \mnras, 432, 1796,
  \dodoi{10.1093/mnras/sts299}

\bibitem[{{Allende Prieto} {et~al.}(2001){Allende Prieto}, {Lambert}, \&
  {Asplund}}]{Allende+Prieto+01}
{Allende Prieto}, C., {Lambert}, D.~L., \& {Asplund}, M. 2001, \apjl, 556, L63,
  \dodoi{10.1086/322874}

\bibitem[{{Barr} {et~al.}(2007){Barr}, {Bedregal}, {Arag{\'o}n-Salamanca},
  {Merrifield}, \& {Bamford}}]{Barr+07}
{Barr}, J.~M., {Bedregal}, A.~G., {Arag{\'o}n-Salamanca}, A., {Merrifield},
  M.~R., \& {Bamford}, S.~P. 2007, \aap, 470, 173,
  \dodoi{10.1051/0004-6361:20077151}

\bibitem[{{Baug} {et~al.}(2019){Baug}, {de Grijs}, {Dewangan}, {Herczeg},
  {Ojha}, {Wang}, {Deng}, \& {Bhatt}}]{Baug+19}
{Baug}, T., {de Grijs}, R., {Dewangan}, L.~K., {et~al.} 2019, \apj, 885, 68,
  \dodoi{10.3847/1538-4357/ab46be}

\bibitem[{{Bekki} \& {Couch}(2011)}]{Bekki+11}
{Bekki}, K., \& {Couch}, W.~J. 2011, \mnras, 415, 1783,
  \dodoi{10.1111/j.1365-2966.2011.18821.x}

\bibitem[{{Berg} {et~al.}(2011){Berg}, {Skillman}, \& {Marble}}]{Berg+11}
{Berg}, D.~A., {Skillman}, E.~D., \& {Marble}, A.~R. 2011, \apj, 738, 2,
  \dodoi{10.1088/0004-637X/738/1/2}

\bibitem[{{Blanton} {et~al.}(2011){Blanton}, {Kazin}, {Muna}, {Weaver}, \&
  {Price-Whelan}}]{Blanton+11}
{Blanton}, M.~R., {Kazin}, E., {Muna}, D., {Weaver}, B.~A., \& {Price-Whelan},
  A. 2011, \aj, 142, 31, \dodoi{10.1088/0004-6256/142/1/31}

\bibitem[{{Bolatto} {et~al.}(2013){Bolatto}, {Wolfire}, \&
  {Leroy}}]{Bolatto+13}
{Bolatto}, A.~D., {Wolfire}, M., \& {Leroy}, A.~K. 2013, \araa, 51, 207,
  \dodoi{10.1146/annurev-astro-082812-140944}

\bibitem[{{Calzetti} {et~al.}(2000){Calzetti}, {Armus}, {Bohlin}, {Kinney},
  {Koornneef}, \& {Storchi-Bergmann}}]{Calzetti+2000}
{Calzetti}, D., {Armus}, L., {Bohlin}, R.~C., {et~al.} 2000, \apj, 533, 682,
  \dodoi{10.1086/308692}

\bibitem[{{Cano-D{\'\i}az} {et~al.}(2022){Cano-D{\'\i}az},
  {Hern{\'a}ndez-Toledo}, {Rodr{\'\i}guez-Puebla}, {Ibarra-Medel},
  {{\'A}vila-Reese}, {Valenzuela}, {Medellin-Hurtado}, {V{\'a}zquez-Mata},
  {Weijmans}, {Gonz{\'a}lez}, {Aquino-Ortiz}, {Mart{\'\i}nez-V{\'a}zquez}, \&
  {Lane}}]{Cano-Diaz+22}
{Cano-D{\'\i}az}, M., {Hern{\'a}ndez-Toledo}, H.~M., {Rodr{\'\i}guez-Puebla},
  A., {et~al.} 2022, \aj, 164, 127, \dodoi{10.3847/1538-3881/ac8549}

\bibitem[{{Cappellari}(2017)}]{Cappellari+17}
{Cappellari}, M. 2017, \mnras, 466, 798, \dodoi{10.1093/mnras/stw3020}

\bibitem[{{Cappellari} \& {Emsellem}(2004)}]{Cappellari+Emsellem+04}
{Cappellari}, M., \& {Emsellem}, E. 2004, \pasp, 116, 138,
  \dodoi{10.1086/381875}

\bibitem[{{Cayrel}(1988)}]{Cayrel+1988}
{Cayrel}, R. 1988, in The Impact of Very High S/N Spectroscopy on Stellar
  Physics, ed. G.~{Cayrel de Strobel} \& M.~{Spite}, Vol. 132, 345

\bibitem[{{Chabrier}(2003)}]{Chabrier+03}
{Chabrier}, G. 2003, \pasp, 115, 763, \dodoi{10.1086/376392}

\bibitem[{{Chen} {et~al.}(2021){Chen}, {Gu}, {Garc{\'i}a-Benito}, {Zhang},
  {Ge}, {Xiao}, \& {Yu}}]{Chen+21}
{Chen}, Z., {Gu}, Q.-S., {Garc{\'i}a-Benito}, R., {et~al.} 2021, \apj, 915, 1,
  \dodoi{10.3847/1538-4357/abfb62}

\bibitem[{{Coccato} {et~al.}(2022){Coccato}, {Fraser-McKelvie}, {Jaff{\'e}},
  {Johnston}, {Cortesi}, \& {Pallero}}]{Coccato+22}
{Coccato}, L., {Fraser-McKelvie}, A., {Jaff{\'e}}, Y.~L., {et~al.} 2022,
  \mnras, 515, 201, \dodoi{10.1093/mnras/stac1764}

\bibitem[{{Copetti} {et~al.}(1986){Copetti}, {Pastoriza}, \&
  {Dottori}}]{Copetti+1986}
{Copetti}, M.~V.~F., {Pastoriza}, M.~G., \& {Dottori}, H.~A. 1986, \aap, 156,
  111

\bibitem[{{Courtois} \& {Tully}(2015)}]{Courtois+15}
{Courtois}, H.~M., \& {Tully}, R.~B. 2015, \mnras, 447, 1531,
  \dodoi{10.1093/mnras/stu2405}

\bibitem[{{Crowther}(2007)}]{Crowther+07}
{Crowther}, P.~A. 2007, \araa, 45, 177,
  \dodoi{10.1146/annurev.astro.45.051806.110615}

\bibitem[{{Crowther} \& {Hadfield}(2006)}]{Crowther+Hadfield+06}
{Crowther}, P.~A., \& {Hadfield}, L.~J. 2006, \aap, 449, 711,
  \dodoi{10.1051/0004-6361:20054298}

\bibitem[{{Cuisinier} {et~al.}(2006){Cuisinier}, {Westera}, {Telles}, \&
  {Buser}}]{Cuisinier+06}
{Cuisinier}, F., {Westera}, P., {Telles}, E., \& {Buser}, R. 2006, \aap, 455,
  825, \dodoi{10.1051/0004-6361:20054637}

\bibitem[{{Curti} {et~al.}(2020){Curti}, {Mannucci}, {Cresci}, \&
  {Maiolino}}]{Curti+20}
{Curti}, M., {Mannucci}, F., {Cresci}, G., \& {Maiolino}, R. 2020, \mnras, 491,
  944, \dodoi{10.1093/mnras/stz2910}

\bibitem[{{Dale} {et~al.}(2009){Dale}, {Cohen}, {Johnson}, {Schuster},
  {Calzetti}, {Engelbracht}, {Gil de Paz}, {Kennicutt}, {Lee}, {Begum},
  {Block}, {Dalcanton}, {Funes}, {Gordon}, {Johnson}, {Marble}, {Sakai},
  {Skillman}, {van Zee}, {Walter}, {Weisz}, {Williams}, {Wu}, \&
  {Wu}}]{Dale+09}
{Dale}, D.~A., {Cohen}, S.~A., {Johnson}, L.~C., {et~al.} 2009, \apj, 703, 517,
  \dodoi{10.1088/0004-637X/703/1/517}

\bibitem[{{de Vaucouleurs} {et~al.}(1991){de Vaucouleurs}, {de Vaucouleurs},
  {Corwin}, {Buta}, {Paturel}, \& {Fouque}}]{de+Vaucouleurs+1991}
{de Vaucouleurs}, G., {de Vaucouleurs}, A., {Corwin}, Herold~G., J., {et~al.}
  1991, {Third Reference Catalogue of Bright Galaxies}

\bibitem[{{Dey} {et~al.}(2019){Dey}, {Schlegel}, {Lang}, {Blum}, {Burleigh},
  {Fan}, {Findlay}, {Finkbeiner}, {Herrera}, {Juneau}, {Landriau}, {Levi},
  {McGreer}, {Meisner}, {Myers}, {Moustakas}, {Nugent}, {Patej}, {Schlafly},
  {Walker}, {Valdes}, {Weaver}, {Y{\`e}che}, {Zou}, {Zhou}, {Abareshi},
  {Abbott}, {Abolfathi}, {Aguilera}, {Alam}, {Allen}, {Alvarez}, {Annis},
  {Ansarinejad}, {Aubert}, {Beechert}, {Bell}, {BenZvi}, {Beutler}, {Bielby},
  {Bolton}, {Brice{\~n}o}, {Buckley-Geer}, {Butler}, {Calamida}, {Carlberg},
  {Carter}, {Casas}, {Castander}, {Choi}, {Comparat}, {Cukanovaite}, {Delubac},
  {DeVries}, {Dey}, {Dhungana}, {Dickinson}, {Ding}, {Donaldson}, {Duan},
  {Duckworth}, {Eftekharzadeh}, {Eisenstein}, {Etourneau}, {Fagrelius},
  {Farihi}, {Fitzpatrick}, {Font-Ribera}, {Fulmer}, {G{\"a}nsicke},
  {Gaztanaga}, {George}, {Gerdes}, {Gontcho}, {Gorgoni}, {Green}, {Guy},
  {Harmer}, {Hernandez}, {Honscheid}, {Huang}, {James}, {Jannuzi}, {Jiang},
  {Joyce}, {Karcher}, {Karkar}, {Kehoe}, {Kneib}, {Kueter-Young}, {Lan},
  {Lauer}, {Le Guillou}, {Le Van Suu}, {Lee}, {Lesser}, {Perreault Levasseur},
  {Li}, {Mann}, {Marshall}, {Mart{\'\i}nez-V{\'a}zquez}, {Martini}, {du Mas des
  Bourboux}, {McManus}, {Meier}, {M{\'e}nard}, {Metcalfe},
  {Mu{\~n}oz-Guti{\'e}rrez}, {Najita}, {Napier}, {Narayan}, {Newman}, {Nie},
  {Nord}, {Norman}, {Olsen}, {Paat}, {Palanque-Delabrouille}, {Peng},
  {Poppett}, {Poremba}, {Prakash}, {Rabinowitz}, {Raichoor}, {Rezaie},
  {Robertson}, {Roe}, {Ross}, {Ross}, {Rudnick}, {Safonova}, {Saha},
  {S{\'a}nchez}, {Savary}, {Schweiker}, {Scott}, {Seo}, {Shan}, {Silva},
  {Slepian}, {Soto}, {Sprayberry}, {Staten}, {Stillman}, {Stupak}, {Summers},
  {Sien Tie}, {Tirado}, {Vargas-Maga{\~n}a}, {Vivas}, {Wechsler}, {Williams},
  {Yang}, {Yang}, {Yapici}, {Zaritsky}, {Zenteno}, {Zhang}, {Zhang}, {Zhou}, \&
  {Zhou}}]{Dey+19}
{Dey}, A., {Schlegel}, D.~J., {Lang}, D., {et~al.} 2019, \aj, 157, 168,
  \dodoi{10.3847/1538-3881/ab089d}

\bibitem[{{Dom{\'\i}nguez S{\'a}nchez} {et~al.}(2020){Dom{\'\i}nguez
  S{\'a}nchez}, {Bernardi}, {Nikakhtar}, {Margalef-Bentabol}, \&
  {Sheth}}]{Dominguez-Sanchez+20}
{Dom{\'\i}nguez S{\'a}nchez}, H., {Bernardi}, M., {Nikakhtar}, F.,
  {Margalef-Bentabol}, B., \& {Sheth}, R.~K. 2020, \mnras, 495, 2894,
  \dodoi{10.1093/mnras/staa1364}

\bibitem[{{Dottori}(1981)}]{Dottori+1981}
{Dottori}, H.~A. 1981, \apss, 80, 267, \dodoi{10.1007/BF00652928}

\bibitem[{{Elbaz} {et~al.}(2007){Elbaz}, {Daddi}, {Le Borgne}, {Dickinson},
  {Alexander}, {Chary}, {Starck}, {Brandt}, {Kitzbichler}, {MacDonald},
  {Nonino}, {Popesso}, {Stern}, \& {Vanzella}}]{Elbaz+07}
{Elbaz}, D., {Daddi}, E., {Le Borgne}, D., {et~al.} 2007, \aap, 468, 33,
  \dodoi{10.1051/0004-6361:20077525}

\bibitem[{{Fairall}(1980)}]{Fairall+1980}
{Fairall}, A.~P. 1980, \mnras, 191, 391, \dodoi{10.1093/mnras/191.2.391}

\bibitem[{{Fraser-McKelvie} {et~al.}(2018{\natexlab{a}}){Fraser-McKelvie},
  {Arag{\'o}n-Salamanca}, {Merrifield}, {Tabor}, {Bernardi}, {Drory}, {Parikh},
  \& {Argudo-Fern{\'a}ndez}}]{Fraser-McKelvie+18b}
{Fraser-McKelvie}, A., {Arag{\'o}n-Salamanca}, A., {Merrifield}, M., {et~al.}
  2018{\natexlab{a}}, \mnras, 481, 5580, \dodoi{10.1093/mnras/sty2563}

\bibitem[{{Fraser-McKelvie} {et~al.}(2018{\natexlab{b}}){Fraser-McKelvie},
  {Brown}, {Pimbblet}, {Dolley}, \& {Bonne}}]{Fraser-McKelvie+18a}
{Fraser-McKelvie}, A., {Brown}, M. J.~I., {Pimbblet}, K., {Dolley}, T., \&
  {Bonne}, N.~J. 2018{\natexlab{b}}, \mnras, 474, 1909,
  \dodoi{10.1093/mnras/stx2823}

\bibitem[{{Garc{\'\i}a-Benito} {et~al.}(2010){Garc{\'\i}a-Benito}, {D{\'\i}az},
  {H{\"a}gele}, {P{\'e}rez-Montero}, {L{\'o}pez}, {V{\'\i}lchez}, {P{\'e}rez},
  {Terlevich}, {Terlevich}, \& {Rosa-Gonz{\'a}lez}}]{Garcia-Benito+10}
{Garc{\'\i}a-Benito}, R., {D{\'\i}az}, A., {H{\"a}gele}, G.~F., {et~al.} 2010,
  \mnras, 408, 2234, \dodoi{10.1111/j.1365-2966.2010.17269.x}

\bibitem[{{Garc{\'\i}a-Benito} {et~al.}(2015){Garc{\'\i}a-Benito}, {Zibetti},
  {S{\'a}nchez}, {Husemann}, {de Amorim}, {Castillo-Morales}, {Cid Fernandes},
  {Ellis}, {Falc{\'o}n-Barroso}, {Galbany}, {Gil de Paz}, {Gonz{\'a}lez
  Delgado}, {Lacerda}, {L{\'o}pez-Fernandez}, {de Lorenzo-C{\'a}ceres},
  {Lyubenova}, {Marino}, {Mast}, {Mendoza}, {P{\'e}rez}, {Vale Asari},
  {Aguerri}, {Ascasibar}, {Bekerait{\.{e}}}, {Bland-Hawthorn},
  {Barrera-Ballesteros}, {Bomans}, {Cano-D{\'\i}az}, {Catal{\'a}n-Torrecilla},
  {Cortijo}, {Delgado-Inglada}, {Demleitner}, {Dettmar}, {D{\'\i}az},
  {Florido}, {Gallazzi}, {Garc{\'\i}a-Lorenzo}, {Gomes}, {Holmes},
  {Iglesias-P{\'a}ramo}, {Jahnke}, {Kalinova}, {Kehrig}, {Kennicutt},
  {L{\'o}pez-S{\'a}nchez}, {M{\'a}rquez}, {Masegosa}, {Meidt}, {Mendez-Abreu},
  {Moll{\'a}}, {Monreal-Ibero}, {Morisset}, {del Olmo}, {Papaderos},
  {P{\'e}rez}, {Quirrenbach}, {Rosales-Ortega}, {Roth}, {Ruiz-Lara},
  {S{\'a}nchez-Bl{\'a}zquez}, {S{\'a}nchez-Menguiano}, {Singh}, {Spekkens},
  {Stanishev}, {Torres-Papaqui}, {van de Ven}, {Vilchez}, {Walcher}, {Wild},
  {Wisotzki}, {Ziegler}, {Alves}, {Barrado}, {Quintana}, \&
  {Aceituno}}]{Garcia-Benito+15}
{Garc{\'\i}a-Benito}, R., {Zibetti}, S., {S{\'a}nchez}, S.~F., {et~al.} 2015,
  \aap, 576, A135, \dodoi{10.1051/0004-6361/201425080}

\bibitem[{{Ge} {et~al.}(2021){Ge}, {Gu}, {Garc{\'\i}a-Benito}, {Lu}, {Lei}, \&
  {Ding}}]{Ge+21}
{Ge}, X., {Gu}, Q.-S., {Garc{\'\i}a-Benito}, R., {et~al.} 2021, \mnras, 507,
  4262, \dodoi{10.1093/mnras/stab2378}

\bibitem[{{Ge} {et~al.}(2020){Ge}, {Gu}, {Garc{\'i}a-Benito}, {Xiao}, \&
  {Li}}]{Ge+20}
{Ge}, X., {Gu}, Q.-S., {Garc{\'i}a-Benito}, R., {Xiao}, M.-Y., \& {Li}, Z.-N.
  2020, \apj, 889, 132, \dodoi{10.3847/1538-4357/ab65f6}

\bibitem[{{Gil de Paz} {et~al.}(2003){Gil de Paz}, {Madore}, \&
  {Pevunova}}]{Gil+de+Paz+03}
{Gil de Paz}, A., {Madore}, B.~F., \& {Pevunova}, O. 2003, \apjs, 147, 29,
  \dodoi{10.1086/374737}

\bibitem[{{Gildas Team}(2013)}]{Gildas+Team+13}
{Gildas Team}. 2013, {GILDAS: Grenoble Image and Line Data Analysis Software}.
\newblock \doeprint{1305.010}

\bibitem[{{Goodman} {et~al.}(2009){Goodman}, {Rosolowsky}, {Borkin}, {Foster},
  {Halle}, {Kauffmann}, \& {Pineda}}]{Goodman+09}
{Goodman}, A.~A., {Rosolowsky}, E.~W., {Borkin}, M.~A., {et~al.} 2009, \nat,
  457, 63, \dodoi{10.1038/nature07609}

\bibitem[{{Guseva} {et~al.}(2000){Guseva}, {Izotov}, \& {Thuan}}]{Guseva+2000}
{Guseva}, N.~G., {Izotov}, Y.~I., \& {Thuan}, T.~X. 2000, \apj, 531, 776,
  \dodoi{10.1086/308489}

\bibitem[{{Haynes} {et~al.}(2018){Haynes}, {Giovanelli}, {Kent}, {Adams},
  {Balonek}, {Craig}, {Fertig}, {Finn}, {Giovanardi}, {Hallenbeck}, {Hess},
  {Hoffman}, {Huang}, {Jones}, {Koopmann}, {Kornreich}, {Leisman}, {Miller},
  {Moorman}, {O'Connor}, {O'Donoghue}, {Papastergis}, {Troischt}, {Stark}, \&
  {Xiao}}]{Haynes+18}
{Haynes}, M.~P., {Giovanelli}, R., {Kent}, B.~R., {et~al.} 2018, \apj, 861, 49,
  \dodoi{10.3847/1538-4357/aac956}

\bibitem[{{Hubble}(1926)}]{Hubble+1926}
{Hubble}, E.~P. 1926, \apj, 64, 321, \dodoi{10.1086/143018}

\bibitem[{{Husemann} {et~al.}(2013){Husemann}, {Jahnke}, {S{\'a}nchez},
  {Barrado}, {Bekerait{\.{e}}}, {Bomans}, {Castillo-Morales},
  {Catal{\'a}n-Torrecilla}, {Cid Fernandes}, {Falc{\'o}n-Barroso},
  {Garc{\'\i}a-Benito}, {Gonz{\'a}lez Delgado}, {Iglesias-P{\'a}ramo},
  {Johnson}, {Kupko}, {L{\'o}pez-Fernandez}, {Lyubenova}, {Marino}, {Mast},
  {Miskolczi}, {Monreal-Ibero}, {Gil de Paz}, {P{\'e}rez}, {P{\'e}rez},
  {Rosales-Ortega}, {Ruiz-Lara}, {Schilling}, {van de Ven}, {Walcher}, {Alves},
  {de Amorim}, {Backsmann}, {Barrera-Ballesteros}, {Bland-Hawthorn}, {Cortijo},
  {Dettmar}, {Demleitner}, {D{\'\i}az}, {Enke}, {Florido}, {Flores}, {Galbany},
  {Gallazzi}, {Garc{\'\i}a-Lorenzo}, {Gomes}, {Gruel}, {Haines}, {Holmes},
  {Jungwiert}, {Kalinova}, {Kehrig}, {Kennicutt}, {Klar}, {Lehnert},
  {L{\'o}pez-S{\'a}nchez}, {de Lorenzo-C{\'a}ceres}, {M{\'a}rmol-Queralt{\'o}},
  {M{\'a}rquez}, {Mendez-Abreu}, {Moll{\'a}}, {del Olmo}, {Meidt}, {Papaderos},
  {Puschnig}, {Quirrenbach}, {Roth}, {S{\'a}nchez-Bl{\'a}zquez}, {Spekkens},
  {Singh}, {Stanishev}, {Trager}, {Vilchez}, {Wild}, {Wisotzki}, {Zibetti}, \&
  {Ziegler}}]{Husemann+13}
{Husemann}, B., {Jahnke}, K., {S{\'a}nchez}, S.~F., {et~al.} 2013, \aap, 549,
  A87, \dodoi{10.1051/0004-6361/201220582}

\bibitem[{{Kauffmann} {et~al.}(2003){Kauffmann}, {Heckman}, {Tremonti},
  {Brinchmann}, {Charlot}, {White}, {Ridgway}, {Brinkmann}, {Fukugita}, {Hall},
  {Ivezi{\'c}}, {Richards}, \& {Schneider}}]{Kauffmann+03}
{Kauffmann}, G., {Heckman}, T.~M., {Tremonti}, C., {et~al.} 2003, \mnras, 346,
  1055, \dodoi{10.1111/j.1365-2966.2003.07154.x}

\bibitem[{Kelz {et~al.}(2006)Kelz, Verheijen, Roth, Bauer, Becker, Paschke,
  Popow, Sánchez, \& Laux}]{Kelz+06}
Kelz, A., Verheijen, M. A.~W., Roth, M.~M., {et~al.} 2006, Publications of the
  Astronomical Society of the Pacific, 118, 129, \dodoi{10.1086/497455}

\bibitem[{{Kennicutt}(1998)}]{Kennicutt+1998}
{Kennicutt}, Robert~C., J. 1998, \araa, 36, 189,
  \dodoi{10.1146/annurev.astro.36.1.189}

\bibitem[{{Kewley} {et~al.}(2001){Kewley}, {Dopita}, {Sutherland}, {Heisler},
  \& {Trevena}}]{Kewley+01}
{Kewley}, L.~J., {Dopita}, M.~A., {Sutherland}, R.~S., {Heisler}, C.~A., \&
  {Trevena}, J. 2001, \apj, 556, 121, \dodoi{10.1086/321545}

\bibitem[{Laurikainen {et~al.}(2006)Laurikainen, Salo, Buta, Knapen, Speltincx,
  \& Block}]{Laurikainen+06}
Laurikainen, E., Salo, H., Buta, R., {et~al.} 2006, The Astronomical Journal,
  132, 2634, \dodoi{10.1086/508810}

\bibitem[{{Leitherer} \& {Heckman}(1995)}]{Leitherer+Heckman+1995}
{Leitherer}, C., \& {Heckman}, T.~M. 1995, \apjs, 96, 9, \dodoi{10.1086/192112}

\bibitem[{{Leitherer} {et~al.}(1999){Leitherer}, {Schaerer}, {Goldader},
  {Delgado}, {Robert}, {Kune}, {de Mello}, {Devost}, \&
  {Heckman}}]{Leitherer+1999}
{Leitherer}, C., {Schaerer}, D., {Goldader}, J.~D., {et~al.} 1999, \apjs, 123,
  3, \dodoi{10.1086/313233}

\bibitem[{{Li} {et~al.}(2020){Li}, {Wang}, {Wu}, {Ma}, \& {Lin}}]{Li+20}
{Li}, C., {Wang}, H.-C., {Wu}, Y.-W., {Ma}, Y.-H., \& {Lin}, L.-H. 2020,
  Research in Astronomy and Astrophysics, 20, 031,
  \dodoi{10.1088/1674-4527/20/3/31}

\bibitem[{{Liang} {et~al.}(2020){Liang}, {Li}, {Li}, {Yan}, {Mo}, {Zhang},
  {Machuca}, \& {Roman-Lopes}}]{Liang+20}
{Liang}, F.-H., {Li}, C., {Li}, N., {et~al.} 2020, \apj, 896, 121,
  \dodoi{10.3847/1538-4357/ab9596}

\bibitem[{{Liang} {et~al.}(2021){Liang}, {Li}, {Li}, {Zhou}, {Yan}, {Mo}, \&
  {Zhang}}]{Liang+21}
---. 2021, \apj, 923, 120, \dodoi{10.3847/1538-4357/ac2bff}

\bibitem[{{L{\'o}pez-S{\'a}nchez} \& {Esteban}(2010)}]{Lopez-Sanchez+10}
{L{\'o}pez-S{\'a}nchez}, {\'A}.~R., \& {Esteban}, C. 2010, \aap, 516, A104,
  \dodoi{10.1051/0004-6361/200913434}

\bibitem[{{Lu} {et~al.}(2022){Lu}, {Gu}, {Ge}, {Ho}, {Gao}, {Chen}, {Xu},
  {Zhang}, {Shi}, {Yuan}, \& {Bao}}]{Lu+22}
{Lu}, S., {Gu}, Q., {Ge}, X., {et~al.} 2022, \apj, 927, 215,
  \dodoi{10.3847/1538-4357/ac4be1}

\bibitem[{{Luo} {et~al.}(2021){Luo}, {Heckman}, {Hwang}, {Rowlands},
  {S{\'a}nchez-Menguiano}, {Riffel}, {Bizyaev}, {Andrews},
  {Fern{\'a}ndez-Trincado}, {Drory}, {S{\'a}nchez Almeida}, {Maiolino}, {Lane},
  \& {Argudo-Fern{\'a}ndez}}]{Luo+21}
{Luo}, Y., {Heckman}, T., {Hwang}, H.-C., {et~al.} 2021, \apj, 908, 183,
  \dodoi{10.3847/1538-4357/abd1df}

\bibitem[{{Maiolino} \& {Mannucci}(2019)}]{Maiolino+Mannucci+19}
{Maiolino}, R., \& {Mannucci}, F. 2019, \aapr, 27, 3,
  \dodoi{10.1007/s00159-018-0112-2}

\bibitem[{{Markwardt}(2009)}]{Markwardt+09}
{Markwardt}, C.~B. 2009, in Astronomical Society of the Pacific Conference
  Series, Vol. 411, Astronomical Data Analysis Software and Systems XVIII, ed.
  D.~A. {Bohlender}, D.~{Durand}, \& P.~{Dowler}, 251.
\newblock \doarXiv{0902.2850}

\bibitem[{{M{\'e}ndez-Abreu} {et~al.}(2018){M{\'e}ndez-Abreu}, {Aguerri},
  {Falc{\'o}n-Barroso}, {Ruiz-Lara}, {S{\'a}nchez-Menguiano}, {de
  Lorenzo-C{\'a}ceres}, {Costantin}, {Catal{\'a}n-Torrecilla}, {Zhu},
  {S{\'a}nchez-Blazquez}, {Florido}, {Corsini}, {Wild}, {Lyubenova}, {van de
  Ven}, {S{\'a}nchez}, {Bland-Hawthorn}, {Galbany}, {Garc{\'\i}a-Benito},
  {Garc{\'\i}a-Lorenzo}, {Gonz{\'a}lez Delgado}, {L{\'o}pez-S{\'a}nchez},
  {Marino}, {M{\'a}rquez}, {Ziegler}, \& {CALIFA
  Collaboration}}]{Mendez-Abreu+18}
{M{\'e}ndez-Abreu}, J., {Aguerri}, J.~A.~L., {Falc{\'o}n-Barroso}, J., {et~al.}
  2018, \mnras, 474, 1307, \dodoi{10.1093/mnras/stx2804}

\bibitem[{{Miralles-Caballero} {et~al.}(2016){Miralles-Caballero}, {D{\'\i}az},
  {L{\'o}pez-S{\'a}nchez}, {Rosales-Ortega}, {Monreal-Ibero},
  {P{\'e}rez-Montero}, {Kehrig}, {Garc{\'\i}a-Benito}, {S{\'a}nchez},
  {Walcher}, {Galbany}, {Iglesias-P{\'a}ramo}, {V{\'\i}lchez}, {Gonz{\'a}lez
  Delgado}, {van de Ven}, {Barrera-Ballesteros}, {Lyubenova}, {Meidt},
  {Falcon-Barroso}, {Mast}, {Mendoza}, \& {CALIFA
  Collaboration}}]{Miralles-Caballero+16}
{Miralles-Caballero}, D., {D{\'\i}az}, A.~I., {L{\'o}pez-S{\'a}nchez},
  {\'A}.~R., {et~al.} 2016, \aap, 592, A105,
  \dodoi{10.1051/0004-6361/201527179}

\bibitem[{{Mishra} {et~al.}(2017){Mishra}, {Barway}, \&
  {Wadadekar}}]{Mishra+17}
{Mishra}, P.~K., {Barway}, S., \& {Wadadekar}, Y. 2017, \mnras, 472, L89,
  \dodoi{10.1093/mnrasl/slx142}

\bibitem[{{Monreal-Ibero} {et~al.}(2012){Monreal-Ibero}, {Walsh}, \&
  {V{\'\i}lchez}}]{Monreal-Ibero+12}
{Monreal-Ibero}, A., {Walsh}, J.~R., \& {V{\'\i}lchez}, J.~M. 2012, \aap, 544,
  A60, \dodoi{10.1051/0004-6361/201219543}

\bibitem[{{Naim} {et~al.}(1995){Naim}, {Lahav}, {Buta}, {Corwin}, {de
  Vaucouleurs}, {Dressler}, {Huchra}, {van den Bergh}, {Raychaudhury}, {Sodre},
  \& {Storrie-Lombardi}}]{Naim+1995}
{Naim}, A., {Lahav}, O., {Buta}, R.~J., {et~al.} 1995, \mnras, 274, 1107,
  \dodoi{10.1093/mnras/274.4.1107}

\bibitem[{{Paudel} {et~al.}(2023){Paudel}, {Yoon}, {Yoo}, {Smith}, {Chhatkuli},
  {Kumar Bachchan}, {Aryal}, {Adhikari}, {Adhikari}, {Sedain}, {Sheikh},
  {Dhital}, {Giri}, \& {Baral}}]{Paudel+23}
{Paudel}, S., {Yoon}, S.-J., {Yoo}, J., {et~al.} 2023, \apjs, 265, 57,
  \dodoi{10.3847/1538-4365/acbfa7}

\bibitem[{{Peng} {et~al.}(2002){Peng}, {Ho}, {Impey}, \& {Rix}}]{Peng+02}
{Peng}, C.~Y., {Ho}, L.~C., {Impey}, C.~D., \& {Rix}, H.-W. 2002, \aj, 124,
  266, \dodoi{10.1086/340952}

\bibitem[{{P{\'e}rez-Montero} {et~al.}(2011){P{\'e}rez-Montero},
  {V{\'\i}lchez}, {Cedr{\'e}s}, {H{\"a}gele}, {Moll{\'a}}, {Kehrig},
  {D{\'\i}az}, {Garc{\'\i}a-Benito}, \&
  {Mart{\'\i}n-Gord{\'o}n}}]{Perez-Montero+11}
{P{\'e}rez-Montero}, E., {V{\'\i}lchez}, J.~M., {Cedr{\'e}s}, B., {et~al.}
  2011, \aap, 532, A141, \dodoi{10.1051/0004-6361/201116582}

\bibitem[{{Pety}(2005)}]{Pety+05}
{Pety}, J. 2005, in SF2A-2005: Semaine de l'Astrophysique Francaise, ed.
  F.~{Casoli}, T.~{Contini}, J.~M. {Hameury}, \& L.~{Pagani}, 721

\bibitem[{{Roy} {et~al.}(2021){Roy}, {Dopita}, {Krumholz}, {Kewley},
  {Sutherland}, \& {Heger}}]{Roy+21}
{Roy}, A., {Dopita}, M.~A., {Krumholz}, M.~R., {et~al.} 2021, \mnras, 502,
  4359, \dodoi{10.1093/mnras/stab376}

\bibitem[{{Salpeter}(1955)}]{Salpeter+1955}
{Salpeter}, E.~E. 1955, \apj, 121, 161, \dodoi{10.1086/145971}

\bibitem[{{S{\'a}nchez} {et~al.}(2016){S{\'a}nchez}, {Garc{\'\i}a-Benito},
  {Zibetti}, {Walcher}, {Husemann}, {Mendoza}, {Galbany}, {Falc{\'o}n-Barroso},
  {Mast}, {Aceituno}, {Aguerri}, {Alves}, {Amorim}, {Ascasibar},
  {Barrado-Navascues}, {Barrera-Ballesteros}, {Bekerait{\`e}},
  {Bland-Hawthorn}, {Cano D{\'\i}az}, {Cid Fernandes}, {Cavichia}, {Cortijo},
  {Dannerbauer}, {Demleitner}, {D{\'\i}az}, {Dettmar}, {de
  Lorenzo-C{\'a}ceres}, {del Olmo}, {Galazzi}, {Garc{\'\i}a-Lorenzo}, {Gil de
  Paz}, {Gonz{\'a}lez Delgado}, {Holmes}, {Igl{\'e}sias-P{\'a}ramo}, {Kehrig},
  {Kelz}, {Kennicutt}, {Kleemann}, {Lacerda}, {L{\'o}pez Fern{\'a}ndez},
  {L{\'o}pez S{\'a}nchez}, {Lyubenova}, {Marino}, {M{\'a}rquez},
  {Mendez-Abreu}, {Moll{\'a}}, {Monreal-Ibero}, {Ortega Minakata},
  {Torres-Papaqui}, {P{\'e}rez}, {Rosales-Ortega}, {Roth},
  {S{\'a}nchez-Bl{\'a}zquez}, {Schilling}, {Spekkens}, {Vale Asari}, {van den
  Bosch}, {van de Ven}, {Vilchez}, {Wild}, {Wisotzki}, {Y{\i}ld{\i}r{\i}m}, \&
  {Ziegler}}]{Sachez+16}
{S{\'a}nchez}, S.~F., {Garc{\'\i}a-Benito}, R., {Zibetti}, S., {et~al.} 2016,
  \aap, 594, A36, \dodoi{10.1051/0004-6361/201628661}

\bibitem[{{S{\'a}nchez-Bl{\'a}zquez} {et~al.}(2006){S{\'a}nchez-Bl{\'a}zquez},
  {Peletier}, {Jim{\'e}nez-Vicente}, {Cardiel}, {Cenarro},
  {Falc{\'o}n-Barroso}, {Gorgas}, {Selam}, \& {Vazdekis}}]{Sanchez-Blazquez+06}
{S{\'a}nchez-Bl{\'a}zquez}, P., {Peletier}, R.~F., {Jim{\'e}nez-Vicente}, J.,
  {et~al.} 2006, \mnras, 371, 703, \dodoi{10.1111/j.1365-2966.2006.10699.x}

\bibitem[{{Sandage} {et~al.}(1970){Sandage}, {Freeman}, \&
  {Stokes}}]{Sandage+1970}
{Sandage}, A., {Freeman}, K.~C., \& {Stokes}, N.~R. 1970, \apj, 160, 831,
  \dodoi{10.1086/150475}

\bibitem[{{Schaerer} \& {Vacca}(1998)}]{Schaerer+Vacca+1998}
{Schaerer}, D., \& {Vacca}, W.~D. 1998, \apj, 497, 618, \dodoi{10.1086/305487}

\bibitem[{{Solomon} \& {Vanden Bout}(2005)}]{Solomon+05}
{Solomon}, P.~M., \& {Vanden Bout}, P.~A. 2005, \araa, 43, 677,
  \dodoi{10.1146/annurev.astro.43.051804.102221}

\bibitem[{{Terlevich} {et~al.}(2004){Terlevich}, {Silich}, {Rosa-Gonz{\'a}lez},
  \& {Terlevich}}]{Terlevich+04}
{Terlevich}, R., {Silich}, S., {Rosa-Gonz{\'a}lez}, D., \& {Terlevich}, E.
  2004, \mnras, 348, 1191, \dodoi{10.1111/j.1365-2966.2004.07432.x}

\bibitem[{{Tresse} {et~al.}(1999){Tresse}, {Maddox}, {Loveday}, \&
  {Singleton}}]{Tresse+1999}
{Tresse}, L., {Maddox}, S., {Loveday}, J., \& {Singleton}, C. 1999, \mnras,
  310, 262, \dodoi{10.1046/j.1365-8711.1999.02977.x}

\bibitem[{{Vazdekis} {et~al.}(2010){Vazdekis}, {S{\'a}nchez-Bl{\'a}zquez},
  {Falc{\'o}n-Barroso}, {Cenarro}, {Beasley}, {Cardiel}, {Gorgas}, \&
  {Peletier}}]{Vazdekis+10}
{Vazdekis}, A., {S{\'a}nchez-Bl{\'a}zquez}, P., {Falc{\'o}n-Barroso}, J.,
  {et~al.} 2010, \mnras, 404, 1639, \dodoi{10.1111/j.1365-2966.2010.16407.x}

\bibitem[{{Verheijen} {et~al.}(2004){Verheijen}, {Bershady}, {Andersen},
  {Swaters}, {Westfall}, {Kelz}, \& {Roth}}]{Verheijen+04}
{Verheijen}, M.~A.~W., {Bershady}, M.~A., {Andersen}, D.~R., {et~al.} 2004,
  Astronomische Nachrichten, 325, 151, \dodoi{10.1002/asna.200310197}

\bibitem[{Welch {et~al.}(2010)Welch, Sage, \& Young}]{Welch+10}
Welch, G.~A., Sage, L.~J., \& Young, L.~M. 2010, The Astrophysical Journal,
  725, 100, \dodoi{10.1088/0004-637x/725/1/100}

\bibitem[{{Willis} {et~al.}(2004){Willis}, {Crowther}, {Fullerton},
  {Hutchings}, {Sonneborn}, {Brownsberger}, {Massa}, \& {Walborn}}]{Willis+04}
{Willis}, A.~J., {Crowther}, P.~A., {Fullerton}, A.~W., {et~al.} 2004, \apjs,
  154, 651, \dodoi{10.1086/422825}

\bibitem[{{Xiao} {et~al.}(2016){Xiao}, {Gu}, {Chen}, \& {Zhou}}]{Xiao+16}
{Xiao}, M.-Y., {Gu}, Q.-S., {Chen}, Y.-M., \& {Zhou}, L. 2016, \apj, 831, 63,
  \dodoi{10.3847/0004-637X/831/1/63}

\bibitem[{{Xu} {et~al.}(2022){Xu}, {Gu}, {Lu}, {Ge}, {Xiao}, \&
  {Contini}}]{Xu+22}
{Xu}, K., {Gu}, Q., {Lu}, S., {et~al.} 2022, \mnras, 509, 1237,
  \dodoi{10.1093/mnras/stab3013}

\bibitem[{{Zhang} {et~al.}(2007){Zhang}, {Kong}, {Li}, {Zhou}, \&
  {Cheng}}]{Zhang+07}
{Zhang}, W., {Kong}, X., {Li}, C., {Zhou}, H.-Y., \& {Cheng}, F.-Z. 2007, \apj,
  655, 851, \dodoi{10.1086/510231}

\end{thebibliography}
\bibliographystyle{aasjournal}
\end{document}